# Analysis of Two-State Folding Using Parabolic Approximation IV: Non-Arrhenius Kinetics of FBP28 WW Part-II

**AUTHOR NAME:** Robert S. Sade

**AUTHOR ADDRESS:** Vinkensteynstraat 128, 2562 TV, Den Haag, Netherlands

**AUTHOR EMAIL ADDRESS:** robert.sade@gmail.com

**AUTHOR AFFILIATION:** Independent Researcher

# ABSTRACT

A model which treats the denatured and the native conformers as being confined to harmonic Gibbs energy wells has been used to rationalize the physical basis for the non-Arrhenius behaviour of spontaneously-folding fixed two-state systems. It is shown that at constant pressure and solvent conditions: (*i*) the rate constant for folding will be a maximum when the heat released upon formation of net molecular interactions is exactly compensated by the heat absorbed to desolvate net polar and non-polar solvent accessible surface area (SASA), as the denatured conformers driven by thermal noise bury their SASA and diffuse on the Gibbs energy surface to reach the activated state; (*ii*) the rate constant for unfolding will be a minimum when the heat absorbed by the native conformers to break various net backbone and sidechain interactions is exactly compensated by the heat of hydration released due to the net increase in SASA, as the native conformers unravel to reach the activated state; (*iii*) the activation entropy for folding will be zero, and the Gibbs barrier to folding will be a minimum, when the decrease in the backbone and the sidechain mobility is exactly compensated by the increase in entropy due to solvent-release, as the denatured conformers bury their SASA to reach the activated state; (*iv*) the activation entropy for unfolding will be zero, and the Gibbs barrier to unfolding will be a maximum when the increase in the backbone and sidechain mobility is exactly compensated by the negentropy of solvent-capture on the protein surface, as the native conformers unravel to reach the activated state; (*v*) while cold denaturation is driven by solvent effects, heat denaturation is primarily due to chain effects; (*vi*) the speed-limit for the folding is ultimately due to conformational searching; and (*vii*) Levinthal's paradox may have little basis if the entropy of solvent-release that accompanies protein folding is taken into consideration.

# INTRODUCTION

It was shown in the preceding papers, henceforth referred to as Papers I-III that the equilibrium and kinetic behaviour of spontaneously-folding fixed two-state systems can be analysed using parabolic approximation. While the theory and assumptions, and the basic equations underlying this procedure were described in Paper-I, equations governing temperature-dependence were derived in Paper-II.[1,2] The framework from Papers I and II was then used to give a detailed description of the non-Arrhenius behaviour of the 37-residue FBP28 WW domain at an unprecedented temperature range and resolution (Paper-III).[3] The purpose of this article is to give a detailed physical explanation for the non-Arrhenius behaviour of two-state systems in terms of *chain* and *solvent effects*, once again, using FBP28 WW as an example.[4] Because this article is primarily an extension of Paper-III, those aspects that were discussed adequately in Paper-III will not be readdressed here; consequently, any critical appraisal of the conclusions drawn here must be done in conjunction with Paper-III in particular, and Papers I and II, in general.

# RESULTS AND DISCUSSION

## Determinants of the enthalpies

### 1. Activation enthalpy for folding

The physical basis for the temperature-dependence of the activation enthalpy for the partial folding reaction $D \rightleftharpoons [TS]$ may be rationalized by deconvoluting it into its formal components.

$$\Delta H_{\text{TS-D}(T)} = \Delta H_{\text{TS-D(res-res)}(T)} + \Delta H_{\text{TS-D(res-solvent)}(T)} + \Delta H_{\text{TS-D(solvent-solvent)}(T)} \qquad (1)$$

$$\Delta H_{\text{TS-D(res-res)}(T)} = \Delta H_{\text{TS-D(backbone)}(T)} + \Delta H_{\text{TS-D(sidechain)}(T)} \qquad (2)$$

$$\Delta H_{\text{TS-D(res-solvent)}(T)} = \Delta H_{\text{TS-D(nonpolar-solvent)}(T)} + \Delta H_{\text{TS-D(polar-solvent)}(T)} \qquad (3)$$

where $\Delta H_{\text{TS-D}(T)}$ is the total change in enthalpy for the activation of conformers from the DSE to the TSE at any given temperature, pressure and solvent conditions, and $\Delta H_{\text{TS-D(res-res)}(T)}$ is purely due to the formation of net backbone and side-chain contacts *en route* to the TSE, i.e., relative to whatever residual structure that pre-exists in the DSE under folding conditions (includes all possible kinds of molecular interactions such as hydrophobic and van der Waals interactions, hydrogen bonding, long-range electrostatic interactions and salt-bridges etc., including the enthalpy of ionization that stems from perturbed $pK_a$s of ionisable residues).[5-8] The term $\Delta H_{TS-D(res-solvent)(T)}$ represents the activation enthalpy due to the desolvation of polar and non-polar residues (changes in the solvation shell; see Fig. 1 in Frauenfelder et al., 2009), while $\Delta H_{TS-D(solvent-solvent)(T)}$ is purely due to the reorganization of the bulk solvent.[9,10]

Since water soluble globular proteins to which this entire discussion is relevant fold in ~55 M of water, to a first approximation, the contribution of the change in enthalpy due to the reorganization of bulk solvent to $\Delta H_{TS-D(T)}$ may be ignored. The reasons are as follows: First, in most *in vitro* protein folding experiments the molar concentration of the protein under investigation ranges from sub-micromolar (in single molecule spectroscopy) to a few hundred micromolar (in NMR studies). Therefore, the ratio of the molar concentration of the bulk water to protein is ~$10^7$ at the lower-end, and about $10^5$ at the higher-end. It is thus not too unreasonable to assume that such an incredibly small amount of solute will not be able to significantly alter the physical properties of bulk water. Second, although the properties of bulk water (density, dielectric constant, surface tension, viscosity etc.) invariably vary with temperature, particularly if the temperature range is substantial, the effects that stem from this variation will cancel out for any given temperature since we are calculating the difference between the values of the state functions of the reaction-states. That one can subtract out the contribution of bulk water is at the heart of differential scanning calorimetry: The heat absorbed or released at constant pressure by the protein+buffer cell is relative to that of the buffer cell. If the thermal behaviour of bulk water in protein+buffer cell is significantly different from the behaviour of water in the buffer cell, then the midpoint of heat denaturation, $T_m$, and the equilibrium enthalpy of unfolding at $T_m$ ($\Delta H_{D-N(cal)(T_m)}$) obtained from calorimetry will not agree with $\Delta H_{D-N(vH)(T_m)}$ (van't Hoff enthalpy) obtained from analysis of a sigmoidal thermal denaturation curve (obtained using spectroscopy, typically CD 217 nm for β-sheet proteins, CD 222 nm for α-helical proteins, and CD 280 nm for tertiary structure) using a two-state approximation (van't Hoff analysis), even if the protein were a legitimate two-state folder.[11,12] If and only if these arguments hold, we may write

$$\Delta H_{TS-D(T)} = \Delta H_{TS-D(backbone+sidechain)(T)} + \Delta H_{TS-D(polar\ solvent+non-polar\ solvent)(T)}$$
$$\equiv \Delta H_{TS-D(chain)(T)} + \Delta H_{TS-D(desolvation)(T)} \quad (4)$$

Of the two terms on the right-hand-side (RHS), the first term due to *chain enthalpy* is negative (exothermic) and favours, while the second term due to *desolvation enthalpy* is positive (endothermic) and disfavours the enthalpic activation of the denatured conformers to the TSE.[9,13] Inspection of **Figure 1** immediately demonstrates that for $T_\alpha \leq T < T_{H(\text{TS-D})}$, the unfavourable desolvation enthalpy dominates over favourable chain enthalpy making it enthalpically unfavourable to activate the conformers in the DSE to the TSE ($\Delta H_{\text{TS-D}(T)} > 0$), and for $T_{H(\text{TS-D})} < T \leq T_\omega$, the favourable chain enthalpy dominates over the unfavourable desolvation enthalpy making it enthalpically favourable to activate the denatured conformers to the TSE ($\Delta H_{\text{TS-D}(T)} < 0$); and these two opposing enthalpies cancel each other out at $T_{H(\text{TS-D})}$ such that $\partial \ln k_{f(T)} / \partial T = \Delta H_{\text{TS-D}(T)} / RT^2 = 0$ and $k_{f(T)}$ is a maximum (**Figure 1−figure supplement 1**). Thus, a corollary is that for a two-state folder at constant pressure and solvent conditions (for example, no change in the pH of the solvent due to the temperature-dependence of the p$K_a$ of the constituent buffer), "*the Gibbs barrier to folding is purely entropic, the solubility of the TSE as compared to the DSE is the greatest, and $k_{f(T)}$ is a maximum, when the heat released upon formation of net molecular interactions is exactly compensated by the heat absorbed to desolvate net polar and non-polar SASA, as the denatured conformers propelled by thermal energy, bury their SASA and diffuse on the Gibbs energy surface to reach the activated state*" (**Figure 1−figure supplement 2**; see Massieu-Planck activation potentials in Paper-III).[3]

We may now take this analysis one step further by introducing the notion of *residual enthalpies*. It is apparent from inspection of Eq. (4) that even if we have no information on the absolute values of the terms on the RHS, if for a given temperature we find that the left-hand-side (LHS) is algebraically positive, it implies that $\Delta H_{\text{TS-D}(T)}$ is purely due to the *residual desolvation enthalpy* ($\Delta H^{\delta+}_{\text{TS-D(desolvation)}(T)} > 0$) which by definition is the positive or endothermic remnant of the algebraic sum of the endothermic desolvation enthalpy and the exothermic chain enthalpy. Conversely, if we find that the LHS is algebraically negative at a given temperature, it implies that $\Delta H_{\text{TS-D}(T)}$ is purely due to the *residual chain enthalpy* ($\Delta H^{\delta-}_{\text{TS-D(chain)}(T)} < 0$) which by definition is the negative or exothermic remnant of the algebraic sum of the endothermic desolvation enthalpy and the exothermic chain enthalpy (the superscripts $\delta^+$ or $\delta^-$ indicate the algebraic sign of the residual quantities). Consequently, we may conclude from inspection of **Figure 1** that for $T_\alpha \leq T < T_{H(\text{TS-D})}$, $\Delta H_{\text{TS-D}(T)}$ is purely due to the *residual desolvation enthalpy* ($\Delta H_{\text{TS-D}(T)} = \Delta H^{\delta+}_{\text{TS-D(desolvation)}(T)} > 0$), making it enthalpically unfavourable; and for $T_{H(\text{TS-D})} < T \leq T_\omega$, $\Delta H_{\text{TS-D}(T)}$ is purely due to the *residual chain enthalpy* ($\Delta H_{\text{TS-D}(T)} = \Delta H^{\delta-}_{\text{TS-D(chain)}(T)} < 0$), making it enthalpically favourable to activate the denatured conformers to the TSE. Naturally, when $T = T_{H(\text{TS-D})}$, the residual enthalpies become zero such that the activation of the denatured conformers to the TSE is enthalpically neutral.

## 2. Change in enthalpy for the partial folding reaction $[TS] \rightleftharpoons N$

Applying similar considerations as above for the second-half of the folding reaction, we may write

$$\begin{aligned}\Delta H_{\text{N-TS}(T)} &= \Delta H_{\text{N-TS(backbone + sidechain)}(T)} + \Delta H_{\text{N-TS(nonpolar-solvent + polar-solvent)}(T)} \\ &\equiv \Delta H_{\text{N-TS(chain)}(T)} + \Delta H_{\text{N-TS(desolvation)}(T)}\end{aligned} \quad (5)$$

Of the two terms on the RHS, the first term due to *chain enthalpy* is negative and favourable, while the second term due to *desolvation enthalpy* is positive and unfavourable. Unlike the $\Delta H_{\text{TS-D}(T)}$ function which changes its algebraic sign only once across the entire temperature range over which a two-state system is physically defined ($T_\alpha \leq T \leq T_\omega$; see Paper-III),[3] the behaviour of $\Delta H_{\text{N-TS}(T)}$ function is far more complex. Inspection of **Figure 2** demonstrates that for the temperature regimes $T_\alpha \leq T < T_{S(\alpha)}$ and $T_{H(\text{TS-N})} < T < T_{S(\omega)}$, the exothermic chain enthalpy dominates over the endothermic desolvation enthalpy. Consequently, we may conclude that the reaction $[TS] \rightleftharpoons N$ is enthalpically favoured and is purely due to the *residual chain enthalpy* ($\Delta H_{\text{N-TS}(T)} = \Delta H^{\delta-}_{\text{N-TS(chain)}(T)} < 0$). In contrast, for the temperature regimes $T_{S(\alpha)} < T < T_{H(\text{TS-N})}$ and $T_{S(\omega)} < T \leq T_\omega$, the endothermic desolvation enthalpy dominates over the exothermic chain enthalpy, leading to $[TS] \rightleftharpoons N$ being enthalpically disfavoured, and is purely due to the *residual desolvation enthalpy* ($\Delta H_{\text{N-TS}(T)} = \Delta H^{\delta+}_{\text{N-TS(desolvation)}(T)} > 0$). At the temperatures $T_{S(\alpha)}$, $T_{H(\text{TS-N})}$, and $T_{S(\omega)}$, the residual enthalpies become zero, such that $[TS] \rightleftharpoons N$ from the viewpoint of enthalpy is neither favoured nor disfavoured ($\Delta H_{\text{N-TS}(T)} = 0$).

## 3. Activation enthalpy for unfolding

If we reverse the reaction-direction, i.e., for the partial unfolding reaction $N \rightleftharpoons [TS]$ (note the change in subscripts that indicate the reaction-direction), we may write

$$\Delta H_{\text{TS-N}(T)} = \Delta H_{\text{TS-N(chain*)}(T)} + \Delta H_{\text{TS-N(hydration)}(T)} \qquad (6)$$

Unlike the first term on the RHS of Eq. (5), $\Delta H_{\text{TS-N(chain*)}(T)}$ on the RHS of Eq. (6) is endothermic since heat is absorbed by the native conformers to break net backbone and side-chain interactions for them to be activated to the TSE. Similarly, unlike the endothermic desolvation enthalpy term on the RHS of Eq. (5), $\Delta H_{\text{TS-N(hydration)}(T)}$ on the RHS of Eq. (6) is exothermic since heat is released upon hydration of polar and non-polar SASA as the native conformers unravel and expose net SASA to reach the TSE. Inspection of **Figure 2−figure supplement 1** demonstrates that for the temperature regimes $T_\alpha \leq T < T_{S(\alpha)}$ and $T_{H(\text{TS-N})} < T < T_{S(\omega)}$, the endothermic $\Delta H_{\text{TS-N(chain*)}(T)}$ term dominates over the exothermic $\Delta H_{\text{TS-N(hydration)}(T)}$ term, such that the activation of the native conformers to the TSE is enthalpically disfavoured, and is purely due to the residual enthalpy that stems from the heat absorbed to break various net backbone and side-chain interactions not being fully compensated by the heat of hydration released due to a net increase in SASA ($\Delta H_{\text{TS-N}(T)} = \Delta H^{\delta+}_{\text{TS-N(chain*)}(T)} > 0$). In contrast, for the temperature regimes $T_{S(\alpha)} < T < T_{H(\text{TS-N})}$ and $T_{S(\omega)} < T \leq T_\omega$, we have $\Delta H_{\text{TS-N(hydration)}(T)} > \Delta H_{\text{TS-N(chain*)}(T)}$; consequently, the activation of the native conformers to the TSE is enthalpically favoured and is purely due to the *residual enthalpy of hydration* ( $\Delta H_{\text{TS-N}(T)} = \Delta H^{\delta-}_{\text{TS-N(hydration)}(T)} < 0$). At $T_{S(\alpha)}$, $T_{H(\text{TS-N})}$, and $T_{S(\omega)}$, we have $\Delta H_{\text{TS-N(chain*)}(T)} = \Delta H_{\text{TS-N(hydration)}(T)}$ such that the activation of the conformers from the NSE to the TSE is neither favoured nor disfavoured ($\Delta H_{\text{TS-N}(T)} = 0$). As explained in Paper-III in considerable detail, although $\partial \ln k_{u(T)} / \partial T = 0 \Rightarrow \Delta H_{\text{TS-N}(T)} = 0$ at $T_{S(\alpha)}$, $T_{H(\text{TS-N})}$, and $T_{S(\omega)}$ (**Figure 2−figure supplement 2**), the behaviour of the system at $T_{S(\alpha)}$ and $T_{S(\omega)}$ is distinctly different from that at $T_{H(\text{TS-N})}$.[3] While $\Delta G_{\text{TS-N}(T)} = \Delta H_{\text{TS-N}(T)} = \Delta S_{\text{TS-N}(T)} = 0$, $\Delta G_{\text{TS-D}(T)} = \Delta G_{\text{N-D}(T)} > 0$, $\Delta S_{\text{TS-D}(T)} = \Delta S_{\text{N-D}(T)} \neq 0$, and $k_{u(T)} = k^0$ (the prefactor in the Arrhenius expression) at $T_{S(\alpha)}$ and $T_{S(\omega)}$, the distinguishing features associated with $T_{H(\text{TS-N})}$ is that $k_{u(T)}$ is a minimum ($k_{u(T)} \ll k^0$), $\Delta G_{\text{TS-N}(T)} > 0$, and $\Delta S_{\text{TS-N}(T)} < 0$ (see activation entropies and Gibbs energies later). Thus, a corollary is that for a two-state folder at constant pressure and solvent conditions, "*the Gibbs barrier to unfolding is purely entropic, the molar concentration of the conformers in the TSE as compared to those in the NSE is the least, and $k_{u(T)}$ is a minimum, when the heat absorbed by the native conformers to break various net backbone and side-chain interactions is exactly compensated by the heat of hydration released due to a net increase in SASA as the native conformers unravel and diffuse on the Gibbs energy surface to reach the TSE*" (**Figure 2−figure supplement 3**).

**4. Change in enthalpy for the coupled reaction $D \rightleftharpoons N$**

Now that the changes in enthalpies for the partial folding reactions $D \rightleftharpoons [TS]$ and $[TS] \rightleftharpoons N$ have been deconvoluted into their constituent chain and desolvation enthalpies across a wide temperature regime, the physical chemistry underlying the variation in $\Delta H_{\text{N-D}(T)}$ (determined independently from thermal denaturation experiments at equilibrium) may be rationalized using the relationship $\Delta H_{\text{N-D}(T)} = \Delta H_{\text{TS-D}(T)} + \Delta H_{\text{N-TS}(T)}$, and by partitioning the physically definable temperature range into six temperature regimes using the reference temperatures $T_\alpha$, $T_{S(\alpha)}$, $T_{H(\text{TS-N})}$, $T_H$, $T_{H(\text{TS-D})}$, $T_{S(\omega)}$, and $T_\omega$ (see **Table 1**).

**Enthalpic Regime I ($T_\alpha \leq T < T_{S(\alpha)}$):** Inspection of **Figure 3** demonstrates that while $D \rightleftharpoons [TS]$ is enthalpically disfavoured and is purely due to the endothermic residual enthalpy of desolvation ($\Delta H^{\delta+}_{\text{TS-D(desolvation)}(T)} > 0$), the reaction $[TS] \rightleftharpoons N$ is enthalpically favourable and is purely due to the exothermic residual chain enthalpy ($\Delta H^{\delta-}_{\text{N-TS(chain)}(T)} < 0$). Because $\Delta H_{\text{N-D}(T)} > 0$ for $T < T_H$ (green curve in **Figure 3B**; see Paper-III and also Becktel and Schellman, 1987),[3,14] we may write

$$\Delta H_{\text{N-D}(T)}\Big|_{T_\alpha \leq T < T_{S(\alpha)}} = \Delta H^{\delta+}_{\text{TS-D(desolvation)}(T)} + \Delta H^{\delta-}_{\text{N-TS(chain)}(T)}\Big|_{T_\alpha \leq T < T_{S(\alpha)}} > 0 \qquad (7)$$

As explained earlier, although we have no information on the absolute values of the terms on the RHS of Eq. (7), we can nevertheless work out from the algebraic sign of the independently determined LHS from thermal denaturation at equilibrium, which one of the terms on the RHS is dominant. Thus, the logical conclusion is that although the second-half of the folding reaction is enthalpically favoured ($\Delta H^{\delta-}_{\text{N-TS(chain)}(T)} < 0$), it is unable to fully compensate for the unfavourable desolvation enthalpy generated in the first-half of the folding reaction ($\Delta H_{\text{TS-D(desolvation)}(T)}^{\delta+} > 0$) such that the coupled reaction $D \rightleftharpoons N$ is enthalpically disfavoured. When $T = T_{S(\alpha)}$, the second term on the RHS becomes zero, leading to $\Delta H_{\text{N-D}(T)}\big|_{T=T_{S(\alpha)}} = \Delta H_{\text{TS-D(desolvation)}(T)}^{\delta+}\big|_{T=T_{S(\alpha)}} > 0$ (intersection of the red and the green curves to the left of the encircled area in **Figure 3B**). This implies that at $T_{S(\alpha)}$, $\Delta H_{\text{N-D}(T)}$ is primarily due to events occurring in the first-half of the folding reaction.

**Enthalpic Regime II ($T_{S(\alpha)} < T < T_{H(\text{TS-N})}$):** Inspection of **Figure 3** shows that for this temperature regime, both $D \rightleftharpoons [TS]$ and $[TS] \rightleftharpoons N$ are enthalpically disfavoured and are purely due to endothermic residual desolvation enthalpies.

$$\Delta H_{\text{N-D}(T)}\big|_{T_{S(\alpha)}<T<T_{H(\text{TS-N})}} = \Delta H_{\text{TS-D(desolvation)}(T)}^{\delta+} + \Delta H_{\text{N-TS(desolvation)}(T)}^{\delta+}\big|_{T_{S(\alpha)}<T<T_{H(\text{TS-N})}} > 0 \qquad (8)$$

Thus, the independently determined endothermic $\Delta H_{\text{N-D}(T)}$ for this regime is once again due to the enthalpic penalty of desolvation, but unlike Regime I, is determined by both the partial folding reactions. When $T = T_{H(\text{TS-N})}$, the second term on the RHS of Eq. (8) becomes zero, leading to $\Delta H_{\text{N-D}(T)}\big|_{T=T_{H(\text{TS-N})}} = \Delta H_{\text{TS-D(desolvation)}(T)}^{\delta+}\big|_{T=T_{H(\text{TS-N})}} > 0$ (intersection of the red and the green curves inside the encircled area in **Figure 3B**). Consequently, we may conclude that at $T_{H(\text{TS-N})}$, $\Delta H_{\text{N-D}(T)}$ is purely due to the endothermic residual desolvation enthalpy incurred in the reaction $D \rightleftharpoons [TS]$. Further, at the two unique temperatures within this regime where $\Delta H_{\text{TS-D}(T)} = \Delta H_{\text{N-TS}(T)}$ (intersection of the blue and the red curves, and indicated by green pointers) we have $\Delta H_{\text{N-D}(T)} = 2\Delta H_{\text{TS-D(desolvation)}(T)}^{\delta+} = 2\Delta H_{\text{N-TS(desolvation)}(T)}^{\delta+} > 0$; and in terms of the absolute enthalpies we have: $H_{\text{TS}(T)} = \left(H_{\text{D}(T)} + H_{\text{N}(T)}\right)/2$.

**Enthalpic Regime III ($T_{H(\text{TS-N})} < T < T_H$):** Inspection of **Figure 3−figure supplement 1** shows that the reaction $D \rightleftharpoons N$ is endothermic and enthalpically disfavoured. What this implies is that although for this temperature regime the reaction $[TS] \rightleftharpoons N$ is enthalpically favoured and is purely due to the exothermic residual chain enthalpy, it nevertheless does not fully compensate for the endothermic residual desolvation enthalpy incurred in the reaction $D \rightleftharpoons [TS]$, such that the coupled reaction $D \rightleftharpoons N$ is enthalpically disfavoured.

$$\Delta H_{\text{N-D}(T)}\big|_{T_{H(\text{TS-N})}<T<T_H} = \Delta H_{\text{TS-D(desolvation)}(T)}^{\delta+} + \Delta H_{\text{N-TS(chain)}(T)}^{\delta-}\big|_{T_{H(\text{TS-N})}<T<T_H} > 0 \qquad (9)$$

When $T = T_H$, we have $\Delta H^{\delta+}_{\text{TS-D(desolvation)}(T)}\big|_{T=T_H} = \Delta H^{\delta-}_{\text{N-TS(chain)}(T)}\big|_{T=T_H}$ such that $\Delta H_{\text{N-D}(T)} = 0$. A corollary is that for a two-state system at constant pressure and solvent conditions, the solubility of the NSE as compared to the DSE, or the equilibrium constant for folding is the greatest, and is driven purely by the difference in entropy between the NSE and the DSE when the endothermic residual desolvation penalty incurred in first-half of the folding reaction is exactly compensated by the exothermic residual chain enthalpy generated in the second-half of the folding reaction.

**Enthalpic Regime IV ($T_H < T < T_{H(\text{TS-D})}$):** Inspection of **Figure 3** and **Figure 3−figure supplement 1** shows that the reaction $D \rightleftharpoons N$ is exothermic and enthalpically favourable ($\Delta H_{\text{N-D}(T)} < 0$). Thus, we may conclude that although the activation of the denatured conformers to the TSE is enthalpically disfavoured and is purely due to the residual endothermic desolvation enthalpy, this is more than compensated by the exothermic residual chain enthalpy generated in the second-half of the folding reaction $[TS] \rightleftharpoons N$, such that the coupled reaction $D \rightleftharpoons N$ is enthalpically favoured.

$$\Delta H_{\text{N-D}(T)}\big|_{T_H < T < T_{H(\text{TS-D})}} = \Delta H^{\delta+}_{\text{TS-D(desolvation)}(T)} + \Delta H^{\delta-}_{\text{N-TS(chain)}(T)}\big|_{T_H < T < T_{H(\text{TS-D})}} < 0 \qquad (10)$$

When $T = T_{H(\text{TS-D})}$, the first term on the RHS of Eq. (10) becomes zero since the chain and desolvation enthalpies for the reaction $D \rightleftharpoons [TS]$ compensate each other exactly. Consequently, we have $\Delta H_{\text{N-D}(T)}\big|_{T=T_{H(\text{TS-D})}} = \Delta H^{\delta-}_{\text{N-TS(chain)}(T)}\big|_{T=T_{H(\text{TS-D})}} < 0$ (intersection of the blue and the green curves in **Figure 3B**), i.e., the exothermic and favourable $\Delta H_{\text{N-D}(T)}$ is primarily due to events occurring in the second-half of the folding reaction.

**Enthalpic Regime V ($T_{H(\text{TS-D})} < T < T_{S(\omega)}$):** It is immediately apparent from inspection of **Figure 3** that the enthalpically favourable coupled reaction $D \rightleftharpoons N$ is a consequence of both the partial folding reactions being enthalpically favourable, and is purely due to exothermic residual chain enthalpy.

$$\Delta H_{\text{N-D}(T)}\big|_{T_{H(\text{TS-D})} < T < T_{S(\omega)}} = \Delta H^{\delta-}_{\text{TS-D(chain)}(T)} + \Delta H^{\delta-}_{\text{N-TS(chain)}(T)}\big|_{T_{H(\text{TS-D})} < T < T_{S(\omega)}} < 0 \qquad (11)$$

When $T = T_{S(\omega)}$, the second term on the RHS of Eq. (11) becomes zero owing to the chain and desolvation enthalpies for the reaction $[TS] \rightleftharpoons N$ compensating each other exactly.

Consequently, we have $\Delta H_{\text{N-D}(T)}\big|_{T=T_{S(\omega)}} = \Delta H^{\delta-}_{\text{TS-D(chain)}(T)}\big|_{T=T_{S(\omega)}} < 0$ (intersection of the red and the green curves to the right of the encircled area in **Figure 3B**), i.e., $\Delta H_{\text{N-D}(T)}$ is primarily due to events occurring in the first-half of the folding reaction. Further, at the temperature where $\Delta H_{\text{TS-D}(T)} = \Delta H_{\text{N-TS}(T)}$ (intersection of the blue and the red curves to the right of the encircled area in **Figure 3B**) we have $\Delta H_{\text{N-D}(T)} = 2\Delta H^{\delta-}_{\text{TS-D(chain)}(T)} = 2\Delta H^{\delta-}_{\text{N-TS(chain)}(T)} < 0$ and $H_{\text{TS}(T)} = \left(H_{\text{D}(T)} + H_{\text{N}(T)}\right)/2$.

**Enthalpic Regime VI ($T_{S(\omega)} < T \leq T_\omega$):** Inspection of **Figure 3** shows that for this regime, the reaction $D \rightleftharpoons [TS]$ which is enthalpically favourable and purely due to the exothermic residual chain enthalpy, more than compensates for the endothermic residual desolvation enthalpy for the reaction $[TS] \rightleftharpoons N$, such that the coupled reaction $D \rightleftharpoons N$ is enthalpically favourable.

$$\Delta H_{\text{N-D}(T)}\big|_{T_{S(\omega)}<T\leq T_\omega} = \Delta H^{\delta-}_{\text{TS-D(chain)}(T)} + \Delta H^{\delta+}_{\text{N-TS(desolvation)}(T)}\big|_{T_{S(\omega)}<T\leq T_\omega} < 0 \qquad (12)$$

## Determinants of entropies

### 1. Activation entropy for folding

The physical basis for the temperature-dependence of the activation entropy for the partial folding reaction $D \rightleftharpoons [TS]$ may be similarly rationalized by deconvoluting them into their formal components.

$$\Delta S_{\text{TS-D}(T)} = \Delta S_{\text{TS-D(res-res)}(T)} + \Delta S_{\text{TS-D(res-solvent)}(T)} + \Delta S_{\text{TS-D(solvent-solvent)}(T)} \qquad (13)$$

$$\Delta S_{\text{TS-D(res-res)}(T)} = \Delta S_{\text{TS-D(backbone)}(T)} + \Delta S_{\text{TS-D(sidechain)}(T)} \qquad (14)$$

$$\Delta S_{\text{TS-D(res-solvent)}(T)} = \Delta S_{\text{TS-D(nonpolar-solvent)}(T)} + \Delta S_{\text{TS-D(polar-solvent)}(T)} \qquad (15)$$

where $\Delta S_{\text{TS-D}(T)}$ is the total change in entropy for the activation of the denatured conformers to the TSE, $\Delta S_{\text{TS-D(res-res)}(T)}$ is the change in entropy due to a change in the backbone and the side-chain mobility, $\Delta S_{\text{TS-D(res-solvent)}(T)}$ is the change in entropy due to reorganization of the solvent molecules in the solvation shell, and $\Delta S_{\text{TS-D(solvent-solvent)}(T)}$ is the change in entropy of the bulk water. As discussed earlier, if to a first approximation we ignore the change in entropy due to reorganization of the bulk solvent, we may write

$$\Delta S_{\text{TS-D}(T)} = \Delta S_{\text{TS-D(backbone + sidechain)}(T)} + \Delta S_{\text{TS-D(nonpolar-solvent + polar-solvent)}(T)}$$
$$\equiv \Delta S_{\text{TS-D(chain)}(T)} + \Delta S_{\text{TS-D(desolvation)}(T)} \tag{16}$$

Because the activation of denatured conformers to the TSE involves a net decrease in backbone and side-chain mobility as compared to the DSE, the first term on the RHS due to *chain entropy* is negative and opposes folding. In contrast, since solvent molecules are released from the solvation shell into the bulk water as the denatured conformers bury net polar and nonpolar SASA *en route* to the TSE, the *desolvation entropy* term is positive and favours the entropic activation of the denatured conformers to the TSE. Consequently, the magnitude and algebraic sign of $\Delta S_{\text{TS-D}(T)}$ is dependent on the intricate temperature-dependent balance between these two opposing entropies.

Inspection of **Figure 4** immediately demonstrates that for $T_\alpha \leq T < T_S$ the favourable entropy of the release of solvent molecules from the solvation shell dominates over the unfavourable chain entropy making it entropically favourable ($\Delta S_{\text{TS-D}(T)} > 0$), and for $T_S < T \leq T_\omega$ the unfavourable chain entropy dominates over the favourable desolvation entropy making it entropically unfavourable ($\Delta S_{\text{TS-D}(T)} < 0$) to activate denatured conformers to the TSE. These two opposing entropies cancel each other out at $T_S$ such that $\partial \Delta G_{\text{TS-D}(T)}/\partial T = -\Delta S_{\text{TS-D}(T)} = 0$ and $\Delta G_{\text{TS-D}(T)}$ is a minimum. A corollary is that for a two-state folder at constant pressure and solvent conditions, "*the difference in SASA between the DSE and the TSE, the position of the TSE relative to the DSE along the heat capacity and entropic RCs, as well as the Gibbs activation energy for folding are all a minimum when the loss of entropy due to decreased backbone and side-chain mobility is exactly compensated by the entropy gained from solvent-release, as the denatured conformers propelled by thermal energy bury their SASA and diffuse on the Gibbs energy surface to reach the TSE*" ("The principle of least displacement").

We may once again take this analysis to another level by introducing the notion of *residual entropies*. Although we have no information on the absolute values of the terms on the RHS of Eq. (16), if we find that at any given temperature the LHS is algebraically positive, it implies that $\Delta S_{\text{TS-D}(T)}$ is purely due to the *residual desolvation entropy* ($\Delta S^{\delta+}_{\text{TS-D(desolvation)}(T)} > 0$) which by definition is the positive or favourable remnant of the algebraic sum of the positive desolvation entropy and the negative chain entropy. Conversely, if the LHS is negative at any given temperature, it implies that $\Delta S_{\text{TS-D}(T)}$ is purely due to the *residual chain entropy* ($\Delta S^{\delta-}_{\text{TS-D(chain)}(T)} < 0$) which by definition is the negative or unfavourable remnant of the algebraic sum of the positive desolvation entropy and the negative chain entropy. Thus, we may conclude from inspection of **Figure 4** that for $T_\alpha \leq T < T_S$, $\Delta S_{\text{TS-D}(T)}$ is purely due to the *residual desolvation entropy* ($\Delta S_{\text{TS-D}(T)} = \Delta S^{\delta+}_{\text{TS-D(desolvation)}(T)} > 0$) making it entropically favourable; and for $T_S < T \leq T_\omega$, $\Delta S_{\text{TS-D}(T)}$ is purely due to the *residual chain entropy* ($\Delta S_{\text{TS-D}(T)} = \Delta S^{\delta-}_{\text{TS-D(chain)}(T)} < 0$) making it entropically unfavourable to activate denatured conformers to the TSE. Obviously, when $T = T_S$, the residual entropies become zero such that the activation of the denatured conformers to the TSE is entropically neutral.

## 2. Change in entropy for the partial folding reaction $[TS] \rightleftharpoons N$

Similarly, for the second-half of the folding reaction we may write

$$\Delta S_{\text{N-TS}(T)} = \Delta S_{\text{N-TS (chain)}(T)} + \Delta S_{\text{N-TS (desolvation)}(T)} \tag{17}$$

The first term on the RHS due to *chain entropy* is negative and unfavourable, while the second term due to *desolvation entropy* is positive and favourable. However, unlike the $\Delta S_{\text{TS-D}(T)}$ function which changes its polarity only once across the physically definable temperature range, the behaviour of the $\Delta S_{\text{N-TS}(T)}$ function with temperature is far more complex. Inspection of **Figure 5** demonstrates that for $T_\alpha \leq T < T_{S(\alpha)}$ and $T_S < T < T_{S(\omega)}$, the negative and unfavourable chain entropy dominates over the positive and favourable desolvation entropy, such that the reaction $[TS] \rightleftharpoons N$ is entropically disfavoured, and is purely due to the *residual chain entropy* ($\Delta S_{\text{N-TS}(T)} = \Delta S^{\delta-}_{\text{N-TS(chain)}(T)} < 0$). In contrast, for $T_{S(\alpha)} < T < T_S$ and $T_{S(\omega)} < T \leq T_\omega$, the positive desolvation entropy dominates over the negative chain entropy, leading to $[TS] \rightleftharpoons N$ being entropically favoured, and is purely due to the *residual desolvation entropy* ($\Delta S_{\text{N-TS}(T)} = \Delta S^{\delta+}_{\text{N-TS(desolvation)}(T)} > 0$). At the temperatures $T_{S(\alpha)}$, $T_S$, and $T_{S(\omega)}$, the residual entropies become zero, such that $[TS] \rightleftharpoons N$ is entropically neither favoured nor disfavoured, i.e., $\partial \Delta G_{\text{N-TS}(T)}/\partial T = -\Delta S_{\text{N-TS}(T)} = 0$. Importantly, while at $T_{S(\alpha)}$ and $T_{S(\omega)}$ we have $G_{\text{TS}(T)} =$ $G_{N(T)}$, $S_{TS(T)} = S_{N(T)}$ and $\Delta G_{N-TS(T)}$ is a maximum, at $T_S$ we have $G_{TS(T)} \gg G_{N(T)}$, $S_{D(T)} = S_{TS(T)} = S_{N(T)}$ with $\Delta G_{N-TS(T)}$ being a minimum and the most negative (see Gibbs energies later). A corollary is that for a two-state folder at constant pressure and solvent conditions, *"the change in Gibbs energy for the flux of the conformers from the TSE to the NSE is most favourable and purely enthalpic when the loss of backbone and side-chain conformational freedom is exactly compensated by the release of solvent from the solvation shell, as the conformers in the TSE bury their SASA to reach the NSE."* Note that the term "flux" is operationally defined as the "diffusion of the conformers from one reaction-state to the other on the Gibbs energy surface."

## 3. Activation entropy for unfolding

If we now reverse the reaction-direction, i.e., for the partial unfolding reaction $N \rightleftharpoons [TS]$ we may write

$$\Delta S_{TS-N(T)} = \Delta S_{TS-N\text{ (chain*)}(T)} + \Delta S_{TS-N\text{ (hydration)}(T)} \tag{18}$$

Unlike the negative first term on the RHS of Eq. (17), $\Delta S_{TS-N\text{ (chain*)}(T)}$ is positive since the backbone and side-chain mobility of the conformers in the TSE is greater than that of the native conformers. In contrast, unlike the positive second term on the RHS of Eq. (17), $\Delta S_{TS-N\text{ (hydration)}(T)}$ is negative since solvent is captured on the protein surface as the native conformers expose net SASA and diffuse on the Gibbs energy surface to reach the TSE. Inspection of **Figure 5−figure supplement 1** demonstrates that for $T_\alpha \leq T < T_{S(\alpha)}$ and $T_S < T < T_{S(\omega)}$, the positive and favourable $\Delta S_{TS-N\text{ (chain*)}(T)}$ term dominates over the negative and unfavourable $\Delta S_{TS-N\text{ (hydration)}(T)}$ term, such that the activation of the native conformers to the TSE is entropically favoured, and is purely due to the residual entropy that stems from the gain in the backbone and side-chain mobility not being fully compensated by the loss of entropy due to solvent capture on the protein surface ($\Delta S_{TS-N(T)} = \Delta S^{\delta+}_{TS-N(\text{chain*})(T)} > 0$). In contrast, for $T_{S(\alpha)} < T < T_S$ and $T_{S(\omega)} < T \leq T_\omega$, the $\Delta S_{TS-N\text{ (hydration)}(T)}$ term dominates over the $\Delta S_{TS-N\text{ (chain*)}(T)}$ leading to the activation of the native conformers to the TSE being entropically disfavoured, and is purely due to the *residual negentropy* of solvent capture ($\Delta S_{TS-N(T)} = \Delta S^{\delta-}_{TS-N(\text{hydration})(T)} < 0$). At the temperatures $T_{S(\alpha)}$, $T_S$, and $T_{S(\omega)}$, the residual entropies become zero, such that the activation of the native conformers to the TSE is entropically neutral, i.e., $\partial \Delta G_{TS\text{-}N(T)}/\partial T = -\Delta S_{TS\text{-}N(T)} = 0$. Because $\Delta G_{TS\text{-}N(T)} = \Delta H_{TS\text{-}N(T)} = \Delta S_{TS\text{-}N(T)} = 0$ at $T_{S(\alpha)}$ and $T_{S(\omega)}$, unfolding is barrierless and $k_{u(T)}$ is an absolute maximum for that particular solvent and pressure, i.e., $k_{u(T)} = k^0$. Further, while the extrema of $\Delta G_{TS\text{-}N(T)}$ are a minimum at $T_{S(\alpha)}$ and $T_{S(\omega)}$, it is a maximum at $T_S$. Thus, for a two-state folder at constant pressure and solvent conditions, "*the Gibbs barriers to unfolding, depending on the temperature, are the greatest as well as the least when the gain in entropy due to the increased backbone and side-chain mobility is exactly compensated by the loss in entropy of the solvent due to its capture on the net SASA exposed, as the native conformers unravel to reach the TSE.*"

## 4. Change in entropy for the coupled reaction $D \rightleftharpoons N$

The deconvolution of the changes in entropies for the partial folding reactions $D \rightleftharpoons [TS]$ and $[TS] \rightleftharpoons N$ into their constituent chain and desolvation entropies enables the physical basis for the temperature-dependence of $\Delta S_{N\text{-}D(T)}$ (determined independently from thermal denaturation experiments at equilibrium using the relationship $R\left[\partial\left(T \ln K_{N\text{-}D(T)}\right)/\partial T\right] = \Delta S_{N\text{-}D(T)}$ where $K_{N\text{-}D(T)}$ is the equilibrium constant for $D \rightleftharpoons N$) to be rationalized. This is best illuminated by partitioning the physically definable temperature range into four regimes using the reference temperatures $T_\alpha$, $T_{S(\alpha)}$, $T_S$, $T_{S(\omega)}$, and $T_\omega$ (see **Table 1**).

**Entropic Regime I ($T_\alpha \leq T < T_{S(\alpha)}$):** Inspection of **Figure 6** demonstrates that while $D \rightleftharpoons [TS]$ is entropically favoured and is purely due to the residual desolvation entropy ($\Delta S^{\delta+}_{TS\text{-}D(desolvation)(T)} > 0$), the reaction $[TS] \rightleftharpoons N$ is entropically disfavoured and is purely due to the residual chain entropy ($\Delta S^{\delta-}_{N\text{-}TS(chain)(T)} < 0$). Because $\Delta S_{N\text{-}D(T)} > 0$ for $T < T_S$ (green curve in **Figure 6B**; see also Paper-III), we may write

$$\Delta S_{N\text{-}D(T)}\Big|_{T_\alpha \leq T < T_{S(\alpha)}} = \Delta S^{\delta+}_{TS\text{-}D(desolvation)(T)} + \Delta S^{\delta-}_{N\text{-}TS(chain)(T)}\Big|_{T_\alpha \leq T < T_{S(\alpha)}} > 0 \qquad (19)$$

Thus, although the second-half of the folding reaction is entropically unfavourable, the favourable desolvation entropy generated in the first-half more than compensates for it, such that the coupled $D \rightleftharpoons N$ is entropically favoured. When $T = T_{S(\alpha)}$, the second term on the RHS becomes zero, leading to $\Delta S_{N\text{-}D(T)}\Big|_{T=T_{S(\alpha)}} = \Delta S^{\delta+}_{TS\text{-}D(desolvation)(T)}\Big|_{T=T_{S(\alpha)}} > 0$ (intersection of the red and the green curves at the extreme left in **Figure 6B**). This implies that at $T_{S(\alpha)}$, the favourable $\Delta S_{\text{N-D}(T)}$ is primarily due to events occurring in the first-half of the folding reaction.

**Entropic Regime II ($T_{S(\alpha)} < T < T_S$):** Inspection of **Figure 6** demonstrates that for this regime both the partial folding reactions are entropically favourable and are purely due to the residual desolvation entropies. Thus we may write

$$\Delta S_{\text{N-D}(T)}\Big|_{T_{S(\alpha)}<T<T_S} = \Delta S^{\delta+}_{\text{TS-D(desolvation)}(T)} + \Delta S^{\delta+}_{\text{N-TS(desolvation)}(T)}\Big|_{T_{S(\alpha)}<T<T_S} > 0 \tag{20}$$

Thus, the logical conclusion is that although the decrease in backbone and sidechain mobility disfavours $D \rightleftharpoons N$, this is more than compensated by the release of solvent from the solvation shell, as the denatured conformers diffuse on the Gibbs energy surface to reach the NSE. Further, at the temperature where $\Delta S_{\text{TS-D}(T)} = \Delta S_{\text{N-TS}(T)}$ (intersection of the blue and the red curves at the extreme left) we have $\Delta S_{\text{N-D}(T)} = 2\Delta S^{\delta+}_{\text{TS-D(desolvation)}(T)} = 2\Delta S^{\delta+}_{\text{N-TS(desolvation)}(T)} > 0$ and $S_{\text{TS}(T)} = \left(S_{\text{D}(T)} + S_{\text{N}(T)}\right)/2$. When $T = T_S$, both the terms on the RHS become zero leading to the Gibbs energy of folding which is the most negative (or $\Delta G_{\text{D-N}(T)}$ is the greatest) being purely enthalpic. A corollary is that "*equilibrium stability is always the greatest, and the position of the TSE relative to the DSE along the SASA, entropic and heat capacity RCs is always the least, when the decrease in the backbone and sidechain entropy that accompanies folding is exactly compensated by the gain in entropy that stems from solvent-release*" ("The principle of least displacement").

**Entropic Regime III ($T_S < T < T_{S(\omega)}$):** Inspection of **Figure 6** demonstrates that for this regime both the partial folding reactions are entropically unfavourable and are purely due to the residual chain entropy. Thus we may write

$$\Delta S_{\text{N-D}(T)}\Big|_{T_S<T<T_{S(\omega)}} = \Delta S^{\delta-}_{\text{TS-D(chain)}(T)} + \Delta S^{\delta-}_{\text{N-TS(chain)}(T)}\Big|_{T_S<T<T_{S(\omega)}} < 0 \tag{21}$$

Thus, although the release of solvent favours both the partial folding reactions, this does not fully compensate for the unfavourable entropy that stems from a decrease in backbone and sidechain mobility. Consequently, the reaction $D \rightleftharpoons N$ is entropically disfavoured and is purely due to residual chain entropy. When $T = T_{S(\omega)}$, the second term on the RHS becomes zero, leading to $\Delta S_{\text{N-D}(T)}\big|_{T=T_{S(\omega)}} = \Delta S^{\delta-}_{\text{TS-D(chain)}(T)}\big|_{T=T_{S(\omega)}} < 0$ (intersection of the red and the green curves). This implies that at $T_{S(\omega)}$, the unfavourable $\Delta S_{\text{N-D}(T)}$ is primarily due to events occurring in the first-half of the folding reaction. Further, at the temperature where $\Delta S_{\text{TS-D}(T)} = \Delta S_{\text{N-TS}(T)}$ (intersection of the blue and the red curves at the extreme right) we once again have the relationship $\Delta S_{\text{N-D}(T)} = 2\Delta S^{\delta+}_{\text{TS-D(desolvation)}(T)} = 2\Delta S^{\delta+}_{\text{N-TS(desolvation)}(T)} > 0$ and $S_{\text{TS}(T)} = \left(S_{\text{D}(T)} + S_{\text{N}(T)}\right)/2$.

**Entropic Regime IV ($T_{S(\omega)} < T \leq T_\omega$):** Inspection of **Figure 6** demonstrates that while $[TS] \rightleftharpoons N$ is entropically favoured and is purely due to the residual desolvation entropy ($\Delta S^{\delta+}_{\text{N-TS(desolvation)}(T)} > 0$), the reaction $D \rightleftharpoons [TS]$ entropically disfavoured and is purely to the residual chain entropy ($\Delta S^{\delta-}_{\text{TS-D(chain)}(T)} < 0$). Because $\Delta S_{\text{N-D}(T)} < 0$ for $T > T_S$, we may write

$$\Delta S_{\text{N-D}(T)}\big|_{T_{S(\omega)} < T \leq T_\omega} = \Delta S^{\delta-}_{\text{TS-D(chain)}(T)} + \Delta S^{\delta+}_{\text{N-TS(desolvation)}(T)}\big|_{T_{S(\omega)} < T \leq T_\omega} < 0 \qquad (22)$$

Thus, the logical conclusion is that although the second-half of the folding reaction is entropically favourable, it is unable to compensate for the unfavourable chain entropy generated in the first-half, such that the coupled $D \rightleftharpoons N$ is entropically disfavoured.

## Determinants of Gibbs energies

Since the changes in enthalpies and entropies have been deconvoluted into their constituent chain and solvent components, it is relatively straightforward to rationalize the physical chemistry underlying the temperature-dependence of the difference in Gibbs energies between the various reaction-states.

**1. Gibbs activation energy for the partial folding reaction $D \rightleftharpoons [TS]$**

This may be discussed by partitioning the physically meaningful temperature range into three distinct regimes using the reference temperatures $T_\alpha$, $T_S$, $T_{H(\text{TS-D})}$, and $T_\omega$ (**Figure 7 and Figure 7−figure supplement 1**).

**Regime I for $\Delta G_{\text{TS-D}(T)}$ ($T_\alpha \leq T < T_S$):** Because $\Delta G_{\text{TS-D}(T)} > 0$, the logical conclusion is that although the entropic component of $\Delta G_{\text{TS-D}(T)}$ favours the activation of the denatured conformers to the TSE and is purely due to the residual desolvation entropy ( $\Delta S_{\text{TS-D(desolvation)}(T)}^{\delta+} > 0$), it does not fully compensate for the unfavourable change in enthalpy that stems purely from residual desolvation enthalpy ($\Delta H_{\text{TS-D(desolvation)}(T)}^{\delta+} > 0$). Thus, we may write

$$\Delta G_{\text{TS-D}(T)}\big|_{T_\alpha \leq T < T_S} = \Delta H_{\text{TS-D(desolvation)}(T)}^{\delta+} - T\Delta S_{\text{TS-D(desolvation)}(T)}^{\delta+}\big|_{T_\alpha \leq T < T_S} > 0 \qquad (23)$$

Because chain parameters do not feature in Eq. (23), we may conclude that the Gibbs activation barrier to folding for this regime is ultimately due to *solvent effects*. At $T_S$, the chain and desolvation entropies compensate each other exactly leading to $\Delta S_{\text{TS-D(desolvation)}(T)}^{\delta+} = 0$. Consequently, we have $\Delta G_{\text{TS-D}(T)}\big|_{T=T_S} = \Delta H_{\text{TS-D(desolvation)}(T)}^{\delta+}\big|_{T=T_S} > 0$. Therefore, the Arrhenius expression for the rate constant for folding at $T_S$ becomes ($k^0$ is the temperature-invariant prefactor)

$$k_{f(T)}\big|_{T=T_S} = k^0 \exp\left(-\frac{\Delta G_{\text{TS-D}(T)}}{RT}\right)\bigg|_{T=T_S} = k^0 \exp\left(-\frac{\Delta H_{\text{TS-D(desolvation)}(T)}^{\delta+}}{RT}\right)\bigg|_{T=T_S} \qquad (24)$$

In summary, the Gibbs barrier to folding is a minimum and is purely due to the endothermic residual desolvation enthalpy, and occurs precisely at $T_S$. Further, at this temperature, equilibrium stability is a maximum, and the position of the TSE relative to the DSE along the SASA, entropic, and heat capacity RCs, is a minimum.

**Regime II for $\Delta G_{\text{TS-D}(T)}$ ($T_S < T < T_{H(\text{TS-D})}$):** In contrast to *Regime I* where the magnitude and algebraic sign of $\Delta G_{\text{TS-D}(T)}$ is determined by the imbalance between unfavourable and favourable terms, *Regime II* is characterised by the unfavourable and endothermic residual desolvation enthalpy ($\Delta H_{\text{TS-D(desolvation)}(T)}^{\delta+} > 0$), and the negative and unfavourable residual chain entropy ($\Delta S_{\text{TS-D(chain)}(T)}^{\delta-} < 0$), colluding to generate a barrier. Thus, we may write

$$\Delta G_{\text{TS-D}(T)}\big|_{T_S < T < T_{H(\text{TS-D})}} = \Delta H_{\text{TS-D(desolvation)}(T)}^{\delta+} - T\Delta S_{\text{TS-D(chain)}(T)}^{\delta-}\big|_{T_S < T < T_{H(\text{TS-D})}} > 0 \qquad (25)$$

Importantly, unlike *Regime I*, the Gibbs barrier to folding for this regime is due to both *chain* and *solvent effects*. At $T = T_{H(\text{TS-D})}$ the chain and desolvation enthalpies compensate each other exactly leading to $\Delta G_{\text{TS-D}(T)}\big|_{T=T_{H(\text{TS-D})}} = -T\Delta S^{\delta-}_{\text{TS-D(chain)}(T)}\big|_{T=T_{H(\text{TS-D})}} > 0$. Therefore, the expression for $k_{f(T)}$ at $T_{H(\text{TS-D})}$ becomes

$$k_{f(T)}\big|_{T=T_{H(\text{TS-D})}} = k^0 \exp\left(-\frac{\Delta G_{\text{TS-D}(T)}}{RT}\right)\bigg|_{T=T_{H(\text{TS-D})}} = k^0 \exp\left(\frac{\Delta S^{\delta-}_{\text{TS-D(chain)}(T)}}{R}\right)\bigg|_{T=T_{H(\text{TS-D})}} \qquad (26)$$

Because $k_{f(T)}$ for any two-state folder at constant pressure and solvent conditions is a maximum at $T_{H(\text{TS-D})}$, the unfavourable residual chain entropy term in Eq. (26) implies that the *speed-limit* for the folding of any two-state folder for a particular solvent and pressure ultimately boils down to *conformational searching*.[15] We will come back to this when we address Levinthal's paradox.[16]

**Regime III for $\Delta G_{\text{TS-D}(T)}$ ($T_{H(\text{TS-D})} < T \leq T_\omega$):** Akin to *Regime I* and unlike *Regime II*, the magnitude and algebraic sign of $\Delta G_{\text{TS-D}(T)}$ for this regime is once again determined by the imbalance between unfavourable and favourable terms. Although the exothermic residual chain enthalpy favours the activation of the denatured conformers to the TSE ($\Delta H^{\delta-}_{\text{TS-D(chain)}(T)} < 0$), it does not fully compensate for the unfavourable residual chain entropy ($\Delta S^{\delta-}_{\text{TS-D(chain)}(T)} < 0$), such that the Gibbs barrier to folding is positive. Thus, we may write

$$\Delta G_{\text{TS-D}(T)}\big|_{T_{H(\text{TS-D})}<T\leq T_\omega} = \Delta H^{\delta-}_{\text{TS-D(chain)}(T)} - T\Delta S^{\delta-}_{\text{TS-D(chain)}(T)}\big|_{T_{H(\text{TS-D})}<T\leq T_\omega} > 0 \qquad (27)$$

Importantly, while *Regime I* is dominated by *solvent effects*, and *Regime II* by *solvent and chain effects*, the Gibbs barrier to folding for *Regime III* is ultimately due to *chain effects*. Further, although the relative contribution of the solvent and chain parameters to $\Delta G_{\text{TS-D}(T)}$ within any given regime is gradual, the switch-over is abrupt and occurs precisely at $T_S$ and $T_{H(\text{TS-D})}$.

**2. Change in Gibbs energy for the partial folding reaction $[TS] \rightleftharpoons N$**

This may be discussed by partitioning the physically meaningful temperature range into five distinct regimes using the reference temperatures $T_\alpha$, $T_{S(\alpha)}$, $T_{H(\text{TS-N})}$, $T_S$, $T_{S(\omega)}$, and $T_\omega$ (**Figure 8 and Figure 8−figure supplement 1**).

**Regime I for $\Delta G_{\text{N-TS}(T)}$ ($T_\alpha \leq T < T_{S(\alpha)}$):** Because $\Delta G_{\text{N-TS}(T)} < 0$, we may conclude that although the reaction $[TS] \rightleftharpoons N$ is entropically disfavoured and is purely due to the residual chain entropy ($\Delta S^{\delta-}_{\text{N-TS(chain)}(T)} < 0$), it is more than compensated by the exothermic residual chain enthalpy ($\Delta H^{\delta-}_{\text{N-TS(chain)}(T)} < 0$). Thus, we may write

$$\Delta G_{\text{N-TS}(T)}\Big|_{T_\alpha \leq T < T_{S(\alpha)}} = \Delta H^{\delta-}_{\text{N-TS(chain)}(T)} - T\Delta S^{\delta-}_{\text{N-TS(chain)}(T)}\Big|_{T_\alpha \leq T < T_{S(\alpha)}} < 0 \tag{28}$$

When $T = T_{S(\alpha)}$, both the terms on the RHS become zero; consequently, $\Delta G_{\text{N-TS}(T)} = -\Delta G_{\text{TS-N}(T)} = 0$ and $k_{u(T)} = k^0$. Because the solvent parameters do not feature in Eq. (28), we may conclude that this regime is dominated by *chain effects*.

**Regime II for $\Delta G_{\text{N-TS}(T)}$ ($T_{S(\alpha)} < T < T_{H(\text{TS-N})}$):** Although the reaction $[TS] \rightleftharpoons N$ is enthalpically disfavoured and is purely due to the residual desolvation enthalpy ($\Delta H^{\delta+}_{\text{N-TS(desolvation)}(T)} > 0$), it is more than compensated by residual desolvation entropy ($\Delta S^{\delta+}_{\text{N-TS(desolvation)}(T)} > 0$), such that $\Delta G_{\text{N-TS}(T)} < 0$. Therefore, we may write

$$\Delta G_{\text{N-TS}(T)}\Big|_{T_{S(\alpha)} < T < T_{H(\text{TS-N})}} = \Delta H^{\delta+}_{\text{N-TS(desolvation)}(T)} - T\Delta S^{\delta+}_{\text{N-TS(desolvation)}(T)}\Big|_{T_{S(\alpha)} < T < T_{H(\text{TS-N})}} < 0 \tag{29}$$

Since chain parameters do not feature in Eq. (29), the change in Gibbs energy for this regime is ultimately due to *solvent effects*. When $T = T_{H(\text{TS-N})}$, the first term on the RHS becomes zero leading to $\Delta G_{\text{N-TS}(T)}\Big|_{T=T_{H(\text{TS-N})}} = -T\Delta S^{\delta+}_{\text{N-TS(desolvation)}(T)}\Big|_{T=T_{H(\text{TS-N})}} < 0$. Now, if we reverse the reaction-direction (i.e., the partial unfolding reaction $N \rightleftharpoons [TS]$) we have $\Delta G_{\text{TS-N}(T)}\Big|_{T=T_{H(\text{TS-N})}} = -T\Delta S^{\delta-}_{\text{TS-N(hydration)}(T)}\Big|_{T=T_{H(\text{TS-N})}} > 0$ where $\Delta S^{\delta-}_{\text{TS-N(hydration)}(T)}$ is the *residual negentropy* of solvent capture (see activation entropy for unfolding). Consequently, the expression for $k_{u(T)}$ which is a minimum at $T_{H(\text{TS-N})}$ becomes

$$k_{u(T)}\Big|_{T=T_{H(\text{TS-N})}} = k^0 \exp\left(-\frac{\Delta G_{\text{TS-N}(T)}}{RT}\right)\Big|_{T=T_{H(\text{TS-N})}} = k^0 \exp\left(\frac{\Delta S^{\delta-}_{\text{TS-N(hydration)}(T)}}{R}\right)\Big|_{T=T_{H(\text{TS-N})}} \tag{30}$$

Note that at this temperature the solubility of the TSE relative to the NSE is a minimum, or the Massieu-Planck activation potential for unfolding is a maximum. Further, in contrast to the maximum of $k_{f(T)}$ being dominated by chain effects, the minimum of $k_{u(T)}$ is ultimately due to the difference in the size of the solvent shells of the conformers in the NSE and the TSE, including their mobility within the solvent shell.

**Regime III for $\Delta G_{\text{N-TS}(T)}$ ($T_{H(\text{TS-N})} < T < T_S$):** Unlike *Regimes I* and *II* where the favourable change in Gibbs energy is due to the favourable terms more than compensating for the unfavourable terms, this regime is characterised by the favourable and exothermic residual chain enthalpy ($\Delta H^{\delta-}_{\text{N-TS(chain)}(T)} < 0$), and the favourable residual desolvation entropy ($\Delta S^{\delta+}_{\text{N-TS(desolvation)}(T)} > 0$) complementing each other, such that $\Delta G_{\text{N-TS}(T)} < 0$. Therefore, we may write

$$\Delta G_{\text{N-TS}(T)}\Big|_{T_{H(\text{TS-N})}<T<T_S} = \Delta H^{\delta-}_{\text{N-TS(chain)}(T)} - T\Delta S^{\delta+}_{\text{N-TS(desolvation)}(T)}\Big|_{T_{H(\text{TS-N})}<T<T_S} < 0 \qquad (31)$$

Because both chain and solvent parameters feature in Eq. (31), we may conclude that this regime is due to both *chain* and *solvent effects*. When $T = T_S$, the second term on the RHS becomes zero leading to $\Delta G_{\text{N-TS}(T)}\Big|_{T=T_S} = \Delta H^{\delta-}_{\text{N-TS(chain)}(T)}\Big|_{T=T_S} < 0$. Now, if we reverse the reaction-direction (i.e., the partial unfolding reaction $N \rightleftharpoons [TS]$) we have $\Delta G_{\text{TS-N}(T)}\Big|_{T=T_S} = \Delta H^{\delta+}_{\text{TS-N(chain*)}(T)}\Big|_{T=T_S} > 0$ where $\Delta H^{\delta+}_{\text{TS-N(chain*)}(T)}$ is the *residual heat* taken up by the native conformers to break various net backbone and side-chain interactions as they are activated to the TSE (see activation enthalpy for unfolding). Consequently, the expression for $k_{u(T)}$ at $T_S$ is given by

$$k_{u(T)}\Big|_{T=T_S} = k^0 \exp\left(-\frac{\Delta G_{\text{TS-N}(T)}}{RT}\right)\Big|_{T=T_S} = k^0 \exp\left(-\frac{\Delta H^{\delta+}_{\text{TS-N(chain*)}(T)}}{RT}\right)\Big|_{T=T_S} \qquad (32)$$

**Regime IV for $\Delta G_{\text{N-TS}(T)}$ ($T_S < T < T_{S(\omega)}$):** Although the reaction $[TS] \rightleftharpoons N$ is entropically disfavoured and is purely due to the residual chain entropy ($\Delta S^{\delta-}_{\text{N-TS(chain)}(T)} < 0$), this is more than compensated by the exothermic chain enthalpy ($\Delta H^{\delta-}_{\text{N-TS(chain)}(T)} < 0$), such that $\Delta G_{\text{N-TS}(T)} < 0$. Therefore, we may write

$$\Delta G_{\text{N-TS}(T)}\Big|_{T_S<T<T_{S(\omega)}} = \Delta H^{\delta-}_{\text{N-TS(chain)}(T)} - T\Delta S^{\delta-}_{\text{N-TS(chain)}(T)}\Big|_{T_S<T<T_{S(\omega)}} < 0 \qquad (33)$$

When $T = T_{S(\omega)}$, both the terms on the RHS become zero; consequently, $\Delta G_{\text{N-TS}(T)} = -\Delta G_{\text{TS-N}(T)} = 0$ and $k_{u(T)} = k^0$. Because solvent parameters do not feature in Eq. (33), this regime is primarily due to *chain effects*.

**Regime V for $\Delta G_{\text{N-TS}(T)}$ ($T_{S(\omega)} < T \leq T_{\omega}$):** Although the reaction $[TS] \rightleftharpoons N$ is enthalpically disfavoured and is purely due to the residual desolvation enthalpy ($\Delta H^{\delta+}_{\text{N-TS(desolvation)}(T)} > 0$), this is more than compensated by the favourable residual desolvation entropy ($\Delta S^{\delta+}_{\text{N-TS(desolvation)}(T)} > 0$), such that $\Delta G_{\text{N-TS}(T)} < 0$. Thus, we may write

$$\Delta G_{\text{N-TS}(T)}\Big|_{T_{S(\omega)} < T \leq T_{\omega}} = \Delta H^{\delta+}_{\text{N-TS(desolvation)}(T)} - T\Delta S^{\delta+}_{\text{N-TS(desolvation)}(T)}\Big|_{T_{S(\omega)} < T \leq T_{\omega}} < 0 \qquad (34)$$

To summarize, while *Regimes I* and *IV* are dominated by chain effects, *Regimes II* and *V* are primarily due to solvent effects. In contrast, *Regime III* is dominated by both chain and solvent effects. Further, while the enthalpic and entropic components complement each other and favour the reaction $[TS] \rightleftharpoons N$ in *Regime III*, $\Delta G_{\text{N-TS}(T)}$ for all the rest of the temperature regimes is due to the dominance of the favourable terms over the unfavourable components.

### 3. Change in Gibbs energy for the coupled reaction $D \rightleftharpoons N$

Now that we have detailed knowledge of how *residual chain and desolvation enthalpies*, and *residual chain and desolvation entropies* battle for dominance, or sometimes collude to determine the temperature-dependence of the magnitude and algebraic sign of the $\Delta G_{\text{TS-D}(T)}$ and $\Delta G_{\text{N-TS}(T)}$ functions, it is relatively straightforward to provide a physical explanation for the behaviour of two-state systems at equilibrium using $\Delta G_{\text{N-D}(T)} = \Delta H_{\text{N-D}(T)} - T\Delta S_{\text{N-D}(T)}$ for the coupled reaction $D \rightleftharpoons N$. This may best be accomplished by partitioning the physically meaningful temperature range into seven distinct regimes using the reference temperatures $T_\alpha$, $T_{S(\alpha)}$, $T_{H(\text{TS-N})}$, $T_H$, $T_S$, $T_{H(\text{TS-D})}$, $T_{S(\omega)}$, and $T_\omega$. However, before we perform a detailed deconvolution, akin to the treatment given to glycolysis in biochemistry textbooks, it is instructive to think of the reaction $D \rightleftharpoons N$ as a business venture. Because $\Delta G_{\text{TS-D}(T)} > 0$ for all temperatures, we may think of the activation of the denatured conformers to the TSE as the "investment or the preparatory phase." In contrast, since $\Delta G_{\text{N-TS}(T)} < 0$ except for the temperatures $T_{S(\alpha)}$ and $T_{S(\omega)}$, we may think of the second-half of the folding reaction as the "pay-off phase" (**Figure 9**). For the temperature regimes $T_\alpha \leq T < T_c$ ($T_c$ is the midpoint of cold denaturation) and $T_m < T \leq T_\omega$ ($T_m$ is the midpoint of heat denaturation), the revenue generated in the pay-off phase does not fully compensate for the investment, and the company incurs a loss. In contrast, for $T_c < T < T_m$, the company makes a net profit since the revenue generated in the pay-off phase more than compensates for the investment. While the net profit at $T_c$ and $T_m$, is zero, the same will be a maximum at $T_S$ since the investment is the least and the pay-off is the greatest (**Figure 10**). Naturally, at $T_{S(\alpha)}$ and $T_{S(\omega)}$ the returns on the investment is null; consequently, the loss incurred is identical to the investment.

**Regime I for $\Delta G_{N-D(T)}$ ($T_\alpha \leq T < T_{S(\alpha)}$):** Because the independently determined $\Delta G_{N-D(T)} > 0$ for this regime, substituting Eqs. (7) and (19) in the Gibbs equation gives

$$\Delta G_{N-D(T)}\big|_{T_\alpha \leq T < T_{S(\alpha)}} = \left( \begin{bmatrix} \Delta H^{\delta+}_{TS-D(desolvation)(T)} + \Delta H^{\delta-}_{N-TS(chain)(T)} \end{bmatrix} \\ -T\begin{bmatrix} \Delta S^{\delta+}_{TS-D(desolvation)(T)} + \Delta S^{\delta-}_{N-TS(chain)(T)} \end{bmatrix} \right)\Bigg|_{T_\alpha \leq T < T_{S(\alpha)}} > 0 \qquad (35)$$

Although chain parameters appear in Eq. (35), since $\Delta H_{N-D(T)}$ and $\Delta S_{N-D(T)}$ are both independently positive for this regime (**Figures 3B** and **6B**), we may conclude that the unfavourable change in Gibbs energy for this regime is ultimately due to the enthalpic penalty of desolvation paid by system to activate the denatured conformers to the TSE, or in short, this regime is dominated by *solvent effects*. When $T = T_{S(\alpha)}$, both $\Delta H^{\delta-}_{N-TS(chain)(T)}$ and $\Delta S^{\delta-}_{N-TS(chain)(T)}$ become zero (or $\Delta G_{N-TS(T)} = 0$; **Figures 2**, **5** and **8**), leading to

$$\Delta G_{N-D(T)}\big|_{T=T_{S(\alpha)}} = \Delta H^{\delta+}_{TS-D(desolvation)(T)} - T\Delta S^{\delta+}_{TS-D(desolvation)(T)}\big|_{T=T_{S(\alpha)}} = \Delta G_{TS-D(T)}\big|_{T=T_{S(\alpha)}} = \lambda > 0 \quad (36)$$

The parameter λ is the *Marcus reorganization energy* for protein folding, and by definition is the Gibbs energy required to compress the DSE under folding conditions to a state whose SASA is identical to that of the NSE but without the stabilizing native interactions (see Paper-I).

**Regime II for $\Delta G_{N-D(T)}$ ($T_{S(\alpha)} < T < T_{H(TS-N)}$):** The relevant equations for this regime are Eqs. (8) and (20). However, since the cold denaturation temperature, $T_c$, at which $\Delta G_{N-D(T)} = RT \ln \left( k_{u(T)} / k_{f(T)} \right) = 0$ (**Figure 10−figure supplement 1**) falls between $T_{S(\alpha)}$ and $T_{H(TS-N)}$ (**Table I**) we may write

$$\Delta G_{\text{N-D}(T)}\Big|_{T_{S(\alpha)}<T<T_{H(\text{TS-N})}} = \left(\begin{bmatrix}\Delta H^{\delta+}_{\text{TS-D(desolvation)}(T)} + \Delta H^{\delta+}_{\text{N-TS(desolvation)}(T)}\end{bmatrix} \\ -T\begin{bmatrix}\Delta S^{\delta+}_{\text{TS-D(desolvation)}(T)} + \Delta S^{\delta+}_{\text{N-TS(desolvation)}(T)}\end{bmatrix}\right)\Bigg|_{T_{S(\alpha)}<T<T_{H(\text{TS-N})}} \quad (37)$$

$$\Rightarrow \Delta G_{\text{N-D}(T)} = \Delta H^{\delta+}_{\text{N-D(desolvation)}(T)} - T\Delta S^{\delta+}_{\text{N-D(desolvation)}(T)} \begin{cases} >0, & T_{S(\alpha)} < T < T_c \\ = 0, & T = T_c \\ < 0, & T_c < T < T_{H(\text{TS-N})} \end{cases} \quad (38)$$

where $\Delta H^{\delta+}_{\text{N-D(desolvation)}(T)} > 0$ is the endothermic residual desolvation enthalpy, and $\Delta S^{\delta+}_{\text{N-D(desolvation)}(T)} > 0$ is the residual desolvation entropy for the coupled reaction $D \rightleftharpoons N$. Thus, for $T_{S(\alpha)} < T < T_c$ we have $\Delta H^{\delta+}_{\text{N-D(desolvation)}(T)} > T\Delta S^{\delta+}_{\text{N-D(desolvation)}(T)}$, and the net flux of the conformers will be from the NSE to the DSE. In contrast, for $T_c < T < T_{H(\text{TS-N})}$ we have $\Delta H^{\delta+}_{\text{N-D(desolvation)}(T)} < T\Delta S^{\delta+}_{\text{N-D(desolvation)}(T)}$, and the net flux of the conformers will be from the DSE to the NSE. When $T = T_c$, the favourable and unfavourable terms on the RHS of Eq. (38) compensate each other exactly leading to

$$\Delta H^{\delta+}_{\text{N-D(desolvation)}(T_c)} = T\Delta S^{\delta+}_{\text{N-D(desolvation)}(T_c)} \Rightarrow T_c = \left(\frac{\Delta H^{\delta+}_{\text{N-D(desolvation)}(T_c)}}{\Delta S^{\delta+}_{\text{N-D(desolvation)}(T_c)}}\right) \quad (39)$$

Consequently, the flux of the conformers from the DSE to the NSE will be identical to the flux in the reverse direction. Now, when $T = T_{H(\text{TS-N})}$, we have $\Delta H^{\delta+}_{\text{N-TS(desolvation)}(T)}\Big|_{T=T_{H(\text{TS-N})}} = 0$ and $k_{u(T)}$ is a minimum (**Figure 2** and **Figure 2−figure supplement 2**), and Eq. (37) becomes

$$\Delta G_{\text{N-D}(T)}\Big|_{T=T_{H(\text{TS-N})}} = \Delta H^{\delta+}_{\text{TS-D(desolvation)}(T)} - T\Delta S^{\delta+}_{\text{N-D(desolvation)}(T)}\Big|_{T=T_{H(\text{TS-N})}} < 0 \quad (40)$$

Thus, the reason why $\Delta G_{\text{N-D}(T)} < 0$ at $T_{H(\text{TS-N})}$ is that although the system incurs an enthalpic penalty to desolvate the protein surface in the reaction $D \rightleftharpoons [TS]$, this is more than compensated by the entropy of solvent-release that accompanies $D \rightleftharpoons N$. Because the chain parameters do not feature in Eqs. (37) - (40), we may conclude that this regime, including cold denaturation, is dominated by *solvent effects*.

**Regime III for $\Delta G_{\text{N-D}(T)}$ ($T_{H(\text{TS-N})} < T < T_H$):** The relevant equations for this regime are Eqs. (9) and (20). Substituting these in the Gibbs equation gives

$$\Delta G_{\text{N-D}(T)}\Big|_{T_{H(\text{TS-N})}<T<T_H} = \begin{pmatrix} \left[\Delta H^{\delta+}_{\text{TS-D(desolvation)}(T)} + \Delta H^{\delta-}_{\text{N-TS(chain)}(T)}\right] \\ -T\left[\Delta S^{\delta+}_{\text{TS-D(desolvation)}(T)} + \Delta S^{\delta+}_{\text{N-TS(desolvation)}(T)}\right] \end{pmatrix}\Bigg|_{T_{H(\text{TS-N})}<T<T_H} < 0 \quad (41)$$

Although the favourable and exothermic residual chain enthalpy term appears in Eq. (41), since $\Delta H_{\text{N-D}(T)}$ and $\Delta S_{\text{N-D}(T)}$ are both independently positive for this regime (**Figures 3B** and **6B**), we may conclude that the favourable change in Gibbs energy for this regime is ultimately due to the favourable entropy of solvent-release, and thus is dominated by *solvent effects*. When $T = T_H$, we have $\Delta H^{\delta+}_{\text{TS-D(desolvation)}(T)} = \Delta H^{\delta-}_{\text{N-TS(chain)}(T)} \Rightarrow \Delta H_{\text{N-D}(T)} = 0$ (**Figure 3**) leading to

$$\Delta G_{\text{N-D}(T)}\Big|_{T=T_H} = -T\Delta S^{\delta+}_{\text{N-D(desolvation)}(T)}\Big|_{T=T_H} < 0 \quad (42)$$

$$\Rightarrow K_{\text{N-D}(T)}\Big|_{T=T_H} = \exp\left(\frac{\Delta S^{\delta+}_{\text{N-D(desolvation)}(T)}}{R}\right)\Bigg|_{T=T_H} \quad (43)$$

Because $K_{\text{N-D}(T)}$ is a maximum at $T_H$, we may conclude that the solubility of the NSE as compared to the DSE is maximum, and is ultimately determined by favourable residual desolvation entropy that stems from the net decrease in the backbone and side-chain mobility being more than compensated by the entropy of net solvent-release, as the denatured conformers propelled by thermal noise bury their SASA and diffuse on the Gibbs energy surface to reach the NSE.

**Regime IV for $\Delta G_{\text{N-D}(T)}$ ($T_H < T < T_S$):** Although the relevant equations for this regime are identical to those describing the behaviour of the previous regime (Eq. (41)) except for the temperature limits, the interpretation is distinctly different.

$$\Delta G_{\text{N-D}(T)}\Big|_{T_H<T<T_S} = \begin{pmatrix} \left[\Delta H^{\delta+}_{\text{TS-D(desolvation)}(T)} + \Delta H^{\delta-}_{\text{N-TS(chain)}(T)}\right] \\ -T\left[\Delta S^{\delta+}_{\text{TS-D(desolvation)}(T)} + \Delta S^{\delta+}_{\text{N-TS(desolvation)}(T)}\right] \end{pmatrix}\Bigg|_{T_H<T<T_S} < 0 \quad (44)$$

Because $\Delta H_{\text{N-D}(T)} < 0$ and $\Delta S_{\text{N-D}(T)} > 0$ for this regime (**Figures 3B** and **6B**), we may conclude that the net flux of the conformers from the DSE to the NSE for this regime is ultimately due to the favourable entropy of solvent-release for the reaction $D \rightleftharpoons N$ ($\Delta S^{\delta+}_{\text{N-D(desolvation)}(T)} = \Delta S^{\delta+}_{\text{TS-D(desolvation)}(T)} + \Delta S^{\delta+}_{\text{N-TS(desolvation)}(T)} > 0$) and the exothermic residual chain enthalpy for the reaction $[TS] \rightleftharpoons N$ ($\Delta H^{\delta-}_{\text{N-TS(chain)}(T)} < 0$) complementing each other. In short, this regime is dominated by both *chain* and *solvent effects*. When $T = T_S$, we have $\Delta S^{\delta+}_{\text{TS-D(desolvation)}(T)} = \Delta S^{\delta+}_{\text{N-TS(desolvation)}(T)} = 0$ (**Figure 6**) and Eq. (44) becomes

$$\Delta G_{\text{N-D}(T)}\Big|_{T=T_S} = \Delta H^{\delta+}_{\text{TS-D(desolvation)}(T)} + \Delta H^{\delta-}_{\text{N-TS(chain)}(T)}\Big|_{T=T_S} < 0 \tag{45}$$

From Schellman's seminal analysis we know that the stability of a two-state system is the greatest (or $\Delta G_{\text{N-D}(T)}$ is the most negative and a minimum) and is purely enthalpic at $T_S$.[14] Eq. (45) tells us that the magnitude of $\Delta G_{\text{N-D}(T)}$ at $T_S$ is ultimately determined by the exothermic residual chain enthalpy generated in the second-half of the folding reaction. Because $T_S - T_H = \Delta H_{\text{D-N}(T_S)}/\Delta C_{p\text{D-N}} = \Delta G_{\text{D-N}(T_S)}/\Delta C_{p\text{D-N}} > 0$ (see Eq. (10) in Becktel and Schellman, 1987), and the difference in heat capacity between the DSE and the NSE is large and positive ($\Delta C_{p\text{D-N}} > 0$), $T_H$ and $T_S$ will not differ by more than a few Kelvin (**Figure 9B** and **Table I**). Despite this small difference in temperature, we see that while the magnitude of $\Delta G_{\text{N-D}(T)}$ is ultimately down to *solvent effects* at $T_H$, it is primarily due to *chain effects* at $T_S$. Further, while both the partial folding reactions take part in generating these *solvent effects* at $T_H$, the *chain effects* at $T_S$ are primarily due to interactions forming in the second-half of the folding reaction.

**Regime V for $\Delta G_{\text{N-D}(T)}$ ($T_S < T < T_{H(\text{TS-D})}$):** The relevant equations for this regime are Eqs. (10) and (21). Substituting these in the Gibbs equation gives

$$\Delta G_{\text{N-D}(T)}\Big|_{T_S<T<T_{H(\text{TS-D})}} = \left(\begin{matrix}\left[\Delta H^{\delta+}_{\text{TS-D(desolvation)}(T)} + \Delta H^{\delta-}_{\text{N-TS(chain)}(T)}\right] \\ -T\left[\Delta S^{\delta-}_{\text{TS-D(chain)}(T)} + \Delta S^{\delta-}_{\text{N-TS(chain)}(T)}\right]\end{matrix}\right)\Bigg|_{T_S<T<T_{H(\text{TS-D})}} < 0$$

$$= \left(\left[\Delta H^{\delta+}_{\text{TS-D(desolvation)}(T)} + \Delta H^{\delta-}_{\text{N-TS(chain)}(T)}\right] - T\Delta S^{\delta-}_{\text{N-D(chain)}(T)}\right)\Bigg|_{T_S<T<T_{H(\text{TS-D})}} < 0 \tag{46}$$

It is immediately apparent from inspection of the terms on the RHS that the only favourable term is the exothermic residual chain enthalpy ($\Delta H^{\delta-}_{\text{N-TS(chain)}(T)} < 0$) generated in the second-half of the folding reaction. Because $\Delta H_{\text{N-D}(T)}$, $\Delta S_{\text{N-D}(T)}$ and $\Delta G_{\text{N-D}(T)}$ are all independently negative for this regime (**Figures 3B**, **6B** and **10B**), we may conclude that the energetically favoured net flux of the conformers from the DSE to the NSE is primarily due to the residual heat liberated from backbone and side-enthalpy in the second-half of the folding reaction more than compensating for all the unfavourable terms, or in short, this regime is dominated by *chain effects*. When $T = T_{H(\text{TS-D})}$, we have $\Delta H^{\delta+}_{\text{TS-D(desolvation)}(T)}\big|_{T=T_{H(\text{TS-D})}} = 0$ and $k_{f(T)}$ is a maximum (**Figure** 1 and **Figure 1−figure supplement 1**), and Eq. (46) becomes

$$\Delta G_{\text{N-D}(T)}\big|_{T=T_{H(\text{TS-D})}} = \Delta H^{\delta-}_{\text{N-TS(chain)}(T)} - T\Delta S^{\delta-}_{\text{N-D(chain)}(T)}\big|_{T=T_{H(\text{TS-D})}} < 0 \qquad (47)$$

**Regime VI for $\Delta G_{\text{N-D}(T)}$ ($T_{H(\text{TS-D})} < T < T_{S(\omega)}$):** The relevant equations for this regime are Eqs. (11) and (21). However, since $T_m$ at which $\Delta G_{\text{N-D}(T)} = RT\ln\left(k_{u(T)}/k_{f(T)}\right) = 0$ (**Figure 10−figure supplement 1**) falls between $T_{H(\text{TS-D})}$ and $T_{S(\omega)}$ (**Table I**) we may write

$$\Delta G_{\text{N-D}(T)}\big|_{T_{H(\text{TS-D})}<T<T_{S(\omega)}} = \left(\begin{array}{c} \Delta H^{\delta-}_{\text{TS-D(chain)}(T)} + \Delta H^{\delta-}_{\text{N-TS(chain)}(T)} \\ -T\left[\Delta S^{\delta-}_{\text{TS-D(chain)}(T)} + \Delta S^{\delta-}_{\text{N-TS(chain)}(T)}\right] \end{array}\right)\Bigg|_{T_{H(\text{TS-D})}<T<T_{S(\omega)}} \qquad (48)$$

$$\Rightarrow \Delta G_{\text{N-D}(T)} = \Delta H^{\delta-}_{\text{N-D(chain)}(T)} - T\Delta S^{\delta-}_{\text{N-D(chain)}(T)} \begin{cases} <0, & T_{H(\text{TS-D})}<T<T_m \\ =0, & T=T_m \\ >0, & T_m<T<T_{S(\omega)} \end{cases} \qquad (49)$$

where $\Delta H^{\delta-}_{\text{N-D(chain)}(T)} < 0$ is the exothermic residual chain enthalpy, and $\Delta S^{\delta-}_{\text{N-D(chain)}(T)} < 0$ is the unfavourable residual chain entropy for the coupled reaction $D \rightleftharpoons N$. Thus, for $T_{H(\text{TS-D})} < T < T_m$ we have $|\Delta H^{\delta-}_{\text{N-D(chain)}(T)}| > |T\Delta S^{\delta-}_{\text{N-D(chain)}(T)}|$, and the net flux of the conformers will be from the DSE to the NSE. In contrast, for $T_m < T < T_{S(\omega)}$ we have $|\Delta H^{\delta-}_{\text{N-D(chain)}(T)}| < |T\Delta S^{\delta-}_{\text{N-D(chain)}(T)}|$ the net flux of the conformers will be from the NSE to the DSE. When $T = T_m$, the favourable and unfavourable terms on the RHS of Eq. (49) compensate each other exactly leading to

$$\Delta H^{\delta-}_{\text{N-D(chain)}(T_m)} = T\Delta S^{\delta-}_{\text{N-D(chain)}(T_m)} \Rightarrow T_m = \left(\frac{\Delta H^{\delta-}_{\text{N-D(chain)}(T_m)}}{\Delta S^{\delta-}_{\text{N-D(chain)}(T_m)}}\right) \qquad (50)$$

Consequently, the flux of the conformers from the DSE to the NSE will be identical to the flux in the reverse direction. When $T = T_{S(\omega)}$, both $\Delta H^{\delta-}_{\text{N-TS(chain)}(T)}$ and $\Delta S^{\delta-}_{\text{N-TS(chain)}(T)}$ become zero (or $\Delta G_{\text{N-TS}(T)} = 0$; **Figures 2**, **5** and **8**), leading to

$$\Delta G_{\text{N-D}(T)}\big|_{T=T_{S(\omega)}} = \Delta H^{\delta-}_{\text{TS-D(chain)}(T)} - T\Delta S^{\delta-}_{\text{TS-D(chain)}(T)}\big|_{T=T_{S(\omega)}} = \Delta G_{\text{TS-D}(T)}\big|_{T=T_{S(\omega)}} = \lambda > 0 \qquad (51)$$

Because solvent parameters do not feature in Eqs. (48) - (51), we may conclude that this regime is dominated by *chain effects*. Importantly, we note that while cold denaturation is driven predominantly by *solvent effects*, heat denaturation is primarily due to *chain effects*.[17]

**Regime VII for $\Delta G_{\text{N-D}(T)}$ ($T_{S(\omega)} < T \leq T_\omega$):** The relevant equations for this regime are Eqs. (12) and (22). Substituting these in the Gibbs equation gives

$$\Delta G_{\text{N-D}(T)}\Big|_{T_{S(\omega)}<T\leq T_\omega} = \begin{bmatrix}\Delta H^{\delta-}_{\text{TS-D(chain)}(T)} + \Delta H^{\delta+}_{\text{N-TS(desolvation)}(T)}\\ -T\left[\Delta S^{\delta-}_{\text{TS-D(chain)}(T)} + \Delta S^{\delta+}_{\text{N-TS(desolvation)}(T)}\right]\end{bmatrix}\Bigg|_{T_{S(\omega)}<T\leq T_\omega} > 0 \qquad (52)$$

Although the favourable and endothermic residual desolvation terms appear in Eq. (52), since $\Delta H_{\text{N-D}(T)}$ and $\Delta S_{\text{N-D}(T)}$ are both independently negative for this regime (**Figures 3B** and **6B**), we may conclude that the unfavourable change in Gibbs energy for this regime is ultimately due to the unfavourable residual chain entropy generated in the second-half of the folding reaction, or in short, this regime is dominated by *chain effects*.

To summarize, we see that the magnitude and algebraic sign of $\Delta G_{\text{N-D}(T)}$ across a wide temperature range is determined by both solvent and chain parameters: While the first three regimes are dominated by *solvent effects* ($T_\alpha \leq T < T_H$), the last three regimes are dominated by *chain effects* ($T_S < T \leq T_\omega$). In contrast, *Regime IV* whose temperature range ($T_H < T < T_S$) is not more than a few Kelvin and is the most stable region, is dominated by both *solvent* and *chain effects*. Importantly, the changeover from solvent-dominated regimes to chain-and-solvent-dominated regime, followed by chain-dominated regimes is abrupt and occurs precisely at $T_H$ and $T_S$, respectively. Further, the temperature-dependence of the state functions of any given two-state system can be modulated by altering either the chain or solvent properties (see *cis*-acting and *trans*-acting factors in Paper-I); and the change brought forth by altering the chain parameters can, in principle, be negated by altering the solvent properties, and *vice versa*.

However, despite its apparent rigour, this deconvolution is far too simplistic. Since *in vitro* protein folding reactions are almost never carried out in water but in a buffer which sometimes also contain other additives, the shell around the protein, although predominantly water, will invariably contain other species. Consequently, the desolvation enthalpy and entropy terms can vary significantly with the solvent composition even if the primary sequence, pH, temperature and pressure are constant. This is because the enthalpic penalty incurred in removing, for example, a neutral or a charged species from the solvent-exposed surface of the denatured conformer before it can be buried *en route* to the NSE *via* the TSE may not be the same as removing water. Similarly, the entropic benefit that stems from stripping the protein surface depends on the nature of the species (for example, structure-making kosmotropes *vs* structure-breaking chaotropes, see Figure 30.12 in Dill and Bromberg, 2003).[18,19] This is precisely why changing the ionic strength, or adding co-solvents can sometimes significantly alter the rate constants and equilibrium parameters, and is the basis for Hofmeister effects (addressed elsewhere).[20-22] Given that *in vitro* folding itself is so complicated despite having precise knowledge of the experimental variables, one can readily comprehend (despite knowing little) the complexity of the folding problem inside the cell.[23,24]

The coupling of partial exergonic and endergonic reactions such that the coupled total reaction is exergonic is a recurring theme in biology for *driven reactions*. Such coupling can occur in *cis* or in *trans*. While spontaneous reversible folding as detailed above is an example of *cis*-coupling, *trans*-coupling can occur either *via* free diffusion and encounter of two species, or through a common interface (protein allostery). *Trans*-coupling is central to signalling cascades, chaperone and chaperonin-mediated folding, and coupled binding and folding etc. A detailed discussion is beyond the scope of this article and is addressed elsewhere.

**Levinthal's paradox**

Levinthal postulated in 1969 that the number of conformations accessible to even modestly-sized polypeptides in their denatured states is astronomical; consequently, he concluded that they will not be able to fold to their native states purely by a random search of all possible conformations.[16] This particular formulation of the question which has since come to be known as the "Levinthal's paradox" enabled the protein folding problem to be explicitly defined and led to ideas such as pathways to protein folding and the kinetic and thermodynamic control of protein folding.[25-30] Although, there have been various attempts to address this paradox− from a monkey's random attempts to type Hamlet's remark "Methinks it is like a weasel,"− to the use of energy landscape funnels and mean first-passage times which essentially suggest that the key to resolving this search problem can be as simple as applying a reasonable energy bias against locally unfavourable conformations,[31-36] we will instead ask "What is the ratio of the *effective* number of conformations in the DSE and the TSE if we account for the positive desolvation entropy that accompanies activation?" Now, if the total number of conformations accessible to the polypeptide in the DSE and the TSE at constant pressure and solvent conditions are denoted by $\Omega_{\text{DSE(chain)}(T)}$ and $\Omega_{\text{TSE(chain)}(T)}$, respectively, then the molar conformational entropies of the DSE and the TSE are given by the Boltzmann expressions $R \ln \Omega_{\text{DSE(chain)}(T)}$ and $R \ln \Omega_{\text{TSE(chain)}(T)}$, respectively. Thus, Eq. (16) may be recast as

$$\Delta S_{\text{TS-D}(T)} = R \ln \frac{\Omega_{\text{TSE(chain)}(T)}}{\Omega_{\text{DSE(chain)}(T)}} + \Delta S_{\text{TS-D(desolvation)}(T)} \tag{53}$$

The first term on the RHS due to chain entropy is negative, while the second term due to desolvation entropy is positive. Thus, for $T_\alpha \leq T < T_S$, the positive second-term on the RHS dominates, and for $T_S < T \leq T_\omega$, the negative first-term dominates, causing the LHS to be positive and negative, respectively; and these two opposing quantities cancel each other out at $T_S$ leading to $\Delta S_{\text{TS-D}(T)} = 0$ (**Figure 4**). Because $\Delta S_{\text{TS-D}(T)} < 0$ for $T_S < T \leq T_\omega$, and is purely due to the residual chain entropy, we may write

$$\Delta S_{\text{TS-D}(T)}\Big|_{T_S < T \leq T_\omega} = \Delta S^{\delta-}_{\text{TS-D(chain)}(T)}\Big|_{T_S < T \leq T_\omega} = R \ln \frac{\Omega_{\text{TSE(chain)}(T)}}{\Omega_{\text{DSE(chain)}(T)}}\Bigg|^{\text{Effective}}_{T_S < T \leq T_\omega} \tag{54}$$

$$\Rightarrow \frac{\Omega_{\text{DSE(chain)}(T)}}{\Omega_{\text{TSE(chain)}(T)}}\Bigg|^{\text{Effective}}_{T_S < T \leq T_\omega} = \exp\left(-\frac{\Delta S_{\text{TS-D}(T)}}{R}\right)\Bigg|_{T_S < T \leq T_\omega} = \exp\left(-\frac{\Delta S^{\delta-}_{\text{TS-D(chain)}(T)}}{R}\right)\Bigg|_{T_S < T \leq T_\omega} \tag{55}$$

It is imperative to note that the ratio on the RHS of Eq. (54) and the LHS of Eq. (55) is not the ratio of *total accessible conformations*, but rather the ratio of *effective number of accessible conformations*. The temperature-dependence of the effective ratio shown in **Figure 11** emphasizes a very important principle: The ratio of total number of conformations available to the polypeptide in the DSE and in the TSE can be a very large number; however, the favourable entropy of solvent-release, depending on the temperature, will partially, or significantly, or even more-than-compensate for the decrease in the backbone and the sidechain conformational freedom, such that the effective ratio is sufficiently small. This is precisely why *foldable* proteins are able to fold within a finite time when temperature, pressure and solvent conditions favour folding. This compensating effect of solvent entropy is certainly not limited to protein folding, and is often invoked in the explanation for anomalous increases in binding energies and strengths of hydrogen bonds, and rate accelerations in enzymatic reactions.[37,38] As an aside, recasting Eq. (26) in terms of Eq. (55) demonstrates that when $T = T_{H(TS-D)}$, the ratio of effective number of accessible conformations in the DSE to those in the TSE is identical to the ratio of the protein folding prefactor and the rate constant for folding.

$$\left.\frac{k^0}{k_{f(T)}}\right|_{T=T_{H(TS-D)}} = \left.\frac{\Omega_{DSE(chain)(T)}}{\Omega_{TSE(chain)(T)}}\right|^{Effective}_{T=T_{H(TS-D)}} \qquad (56)$$

Thus, from the perspective of the parabolic hypothesis, Levinthal's paradox appears to have little basis for $T_c < T < T_m$, where the NSE is more stable than the DSE (note that we cannot calculate the effective ratio for $T < T_S$ since for this temperature regime the entropy of solvent release that accompanies folding more than compensates for the unfavourable decrease in chain entropy). Indeed, if the effective ratio were astronomically large, as is the case at $T = T_\omega$ ($10^{149}$), the protein ideally will not be able to fold; and the only reason why the folding time is finite at $T = T_\omega$ is that it is partially compensated by the exothermic change in enthalpy ($k_{f(T)} = 0.2815 \, \text{s}^{-1} \Rightarrow \tau = 1/k_{f(T)} = 3.55 \, \text{s}$, **Figures 1A** and **7B**). However, the criticism of the paradox cannot be levelled at Levinthal since he never stated that there exists a paradox;[16,39] and the notion that there exist an astronomical number of conformations in the DSE is for a hypothetical case which may not be relevant to foldable proteins since their DSEs under folding conditions are never extended chains.[7,8,40,41] In fact, Levinthal's explanation offers a powerful insight into what might be happening in real scenarios as is evident from: "*In nature, proteins apparently do not sample all of these possible configurations since they fold in a few seconds, and even postulating a minimum time from one conformation to another, the proteins would have to try on the order of $10^8$ different conformations at most before reaching their final state.*"[16] The unfortunate propagation of this paradox probably has to do with the early work on the DSEs of proteins by Tanford and co-workers, which likened them to random coils in high concentration of denaturants.[42,43] Consequently, the "random coil" approximation for the DSEs of proteins− which essentially implies that all possible conformational states in the DSE have equal probability of being populated− became a deeply entrenched idea until the early nineties for two predominant reasons: (*i*) paucity of high-resolution structural data on the DSEs of proteins under folding conditions that suggested otherwise; and (*ii*) the relative ease of interpretation of the effect of perturbations (mutations, denaturants, co-solvents, temperature, pressure etc.) on folding equilibria, i.e., if the DSEs are assumed to be a random coils, the effect of perturbations on the energetics of the DSEs can be ignored, and conveniently be attributed to the native states whose structure is known to atomic or near-atomic resolution.[7] Although it became increasingly apparent by the early nineties that the DSEs of proteins are not random coils,[6,44,45] the discovery of proteins that fold in a simple two-state manner, and the introduction of two-point Brønsted analysis (Φ-value analysis) shifted the emphasis onto the TSEs of proteins.[46,47] Since we now know that not only do the DSEs under folding conditions have a significant amount of residual structure that includes both short-range and long-range interactions, that they can be native-like, and can persist even under high concentration of denaturants, it is perhaps not too unreasonable to expect that even the total conformational space accessible to the denatured polypeptide under folding conditions may not be astronomical but restricted for the temperature-range within which the NSE is more stable than the DSE.[7,8,48-51]

In conclusion, any explanation for protein folding that focusses purely on chain entropy and underestimates the contribution of the entropy of solvent-release will not only be inadequate but will also be misleading because it will erroneously: (*i*) portray protein folding as a phenomenon that is far more complex and daunting than it actually is; (*ii*) imply through use of the term "information" that the principles that govern protein folding are fundamentally different from those that determine the chemical reactions of small molecules; (*iii*) imply that there is such a thing as a "folding code" that translates a 1-dimensional chain into its 3-dimensional structure; and (*iv*) imply that the protein folding problem was solved by evolution, since in all probability simple polymers would have folded and unfolded for possibly more than a billion years before life as we know it came into being.[52] It is precisely for this reason it is illogical to state that "proteins fold in biologically relevant timescales" because the time-scale of biology is a consequence of physical chemistry, or as Feynman put it so well, *"there is nothing that living things do that cannot be understood from the point of view that they are made of atoms acting according to the laws of physics."*[53,54]

## CONCLUDING REMARKS

Owing to space constraints, the physics behind the origin of extreme thermal stability has not been addressed in this article. Nevertheless, for two proteins of identical chain lengths but differing primary sequences, a decrease in $\Delta C_{p\text{D-N}}$ will concomitantly lead to a decrease in $\Delta C_{p\text{D-TS}(T)}$; and because $\Delta C_{p\text{TS-D}(T)}$ is the slope of the $\Delta H_{\text{TS-D}(T)}$ function, it essentially implies that $\Delta H_{\text{TS-D}(T)}$ becomes relatively insensitive to temperature such that $T_{H(\text{TS-D})}$ is shifted to higher temperatures. Equivalently, since the slope of $\Delta S_{\text{TS-D}(T)}$ is also related to $\Delta C_{p\text{TS-D}(T)}$, it implies that $\Delta S_{\text{TS-D}(T)}$ will also be relatively insensitive to temperature such that $T_S$ is shifted to higher temperatures. Because the reference temperatures are interrelated (**Table 1**), a shift in $T_S$ and $T_{H(\text{TS-D})}$ to higher temperatures implies a concomitant shift in $T_m$ to higher temperatures. Although this can be achieved either by decreasing the SASA of the DSE (i.e., a more compact DSE) or increasing the SASA of the NSE (i.e., a more expanded NSE) or both, if the said two-state systems share the same native fold and similar primary sequences, increasing the SASA of the NSE would be an unlikely scenario; instead, thermal tolerance would primarily stem from changes in the SASA of the DSE. Because the SASA of the DSE is itself a function of its residual structure, all that needs to be done is increase the residual structure in the DSE. This can be achieved *via* introduction of hydrophobic clusters, charge clusters and ion-pairs, disulfide bonds, metal ion coordination to residues such as histidines etc. in the DSE. These effects will be addressed elsewhere.

## COMPETING FINANCIAL INTERESTS

The author declares no competing financial interests.

## COPYRIGHT INFORMATION

# Table 1: Reference temperatures

| Temperature | Value | Remark |
|---|---|---|
| $T_\alpha$ | 182 K | A two-state system is physically undefined for $T < T_\alpha$ |
| $T_{S(\alpha)}$ | 184.4 K | $m_{\text{TS-N}(T)} = 0$, $\Delta H_{\text{TS-N}(T)} = \Delta S_{\text{TS-N}(T)} = \Delta G_{\text{TS-N}(T)} = 0$, $k_{u(T)} = k^0$ |
| $T_{C_p\text{TS-N}(\alpha)}$ | 201 K | $\Delta C_{p\text{TS-N}(T)} = 0$ |
| $T_c$ | 223.6 K | Midpoint of cold denaturation, $\Delta G_{\text{D-N}(T)} = 0$, $k_{f(T)} = k_{u(T)}$ |
| $T_{H(\text{TS-N})}$ | 264.3 K | $\Delta H_{\text{TS-N}(T)} = 0$, $k_{u(T)}$ is a minimum |
| $T_H$ | 272.9 K | $\Delta H_{\text{TS-D}(T)} = \Delta H_{\text{TS-N}(T)}$, $\Delta H_{\text{D-N}(T)} = 0$, $\Delta H_{\text{TS-D}(T)} > 0$, $\Delta H_{\text{TS-N}(T)} > 0$, |
| $T_S$ | 278.8 K | $\Delta S_{\text{TS-D}(T)} = \Delta S_{\text{TS-N}(T)} = \Delta S_{\text{D-N}(T)} = 0$, $\Delta G_{\text{D-N}(T)}$ is a maximum |
| $T_{H(\text{TS-D})}$ | 311.4 K | $\Delta H_{\text{TS-D}(T)} = 0$, $k_{f(T)}$ is a maximum |
| $T_m$ | 337.2 K | Midpoint of heat denaturation, $\Delta G_{\text{D-N}(T)} = 0$, $k_{f(T)} = k_{u(T)}$ |
| $T_{C_p\text{TS-N}(\omega)}$ | 361.7 K | $\Delta C_{p\text{TS-N}(T)} = 0$ |
| $T_{S(\omega)}$ | 384.5 K | $m_{\text{TS-N}(T)} = 0$, $\Delta H_{\text{TS-N}(T)} = \Delta S_{\text{TS-N}(T)} = \Delta G_{\text{TS-N}(T)} = 0$, $k_{u(T)} = k^0$ |
| $T_\omega$ | 388 K | A two-state system is physically undefined for $T > T_\omega$ |

# FIGURES

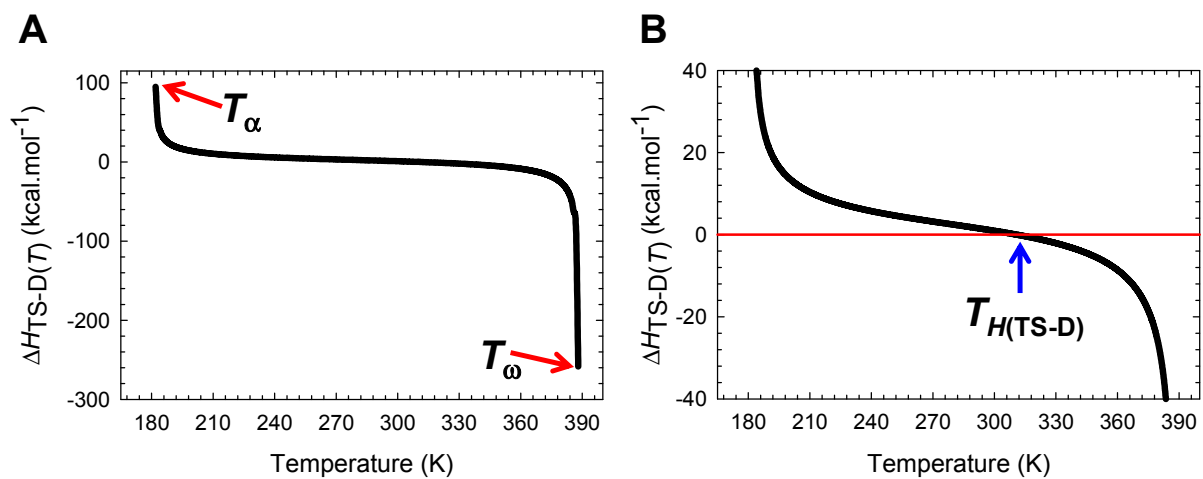

**Figure 1.**

**Temperature-dependence of the activation enthalpy for folding.**

**(A)** The variation in function with temperature. The slope of this curve varies with temperature, equals $\Delta C_{p\text{TS-D}(T)}$, and is algebraically negative. **(B)** An appropriately scaled version of the plot on the left to illuminate the three important scenarios: (*i*) $\Delta H_{\text{TS-D}(T)} > 0$ for $T_\alpha \leq T < T_{H(\text{TS-D})}$; (*ii*) $\Delta H_{\text{TS-D}(T)} < 0$ for $T_{H(\text{TS-D})} < T \leq T_\omega$; and (*iii*) $\Delta H_{\text{TS-D}(T)} = 0$ when $T = T_{H(\text{TS-D})}$. Note that $k_{f(T)}$ is a maximum at $T_{H(\text{TS-D})}$. The values of the reference temperatures are given in **Table 1**.

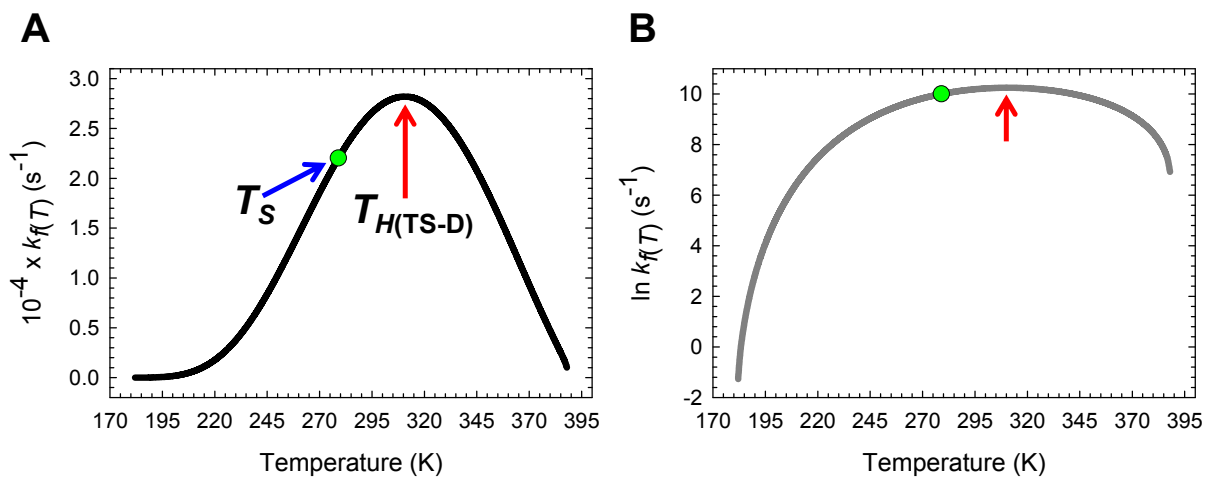

**Figure 1−figure supplement 1.**

**Temperature-dependence of $k_{f(T)}$.**

**(A)** Temperature-dependence of $k_{f(T)}$ on a linear scale. The slope of this curve is given by $k_{f(T)}\Delta H_{TS-D(T)}/RT^2$. **(B)** Temperature-dependence of $k_{f(T)}$ on a logarithmic scale. The slope of this curve is given by $\Delta H_{TS-D(T)}/RT^2$. The green dots represent the temperature $T_S$ at which $\Delta G_{D-N(T)}$ is a maximum, $\Delta G_{TS-D(T)}$ is a minimum, and the absolute entropies of the DSE, the TSE and the NSE are identical. The red pointers indicate the temperature $T_{H(TS-D)}$ at which $k_{f(T)}$ is a maximum, $\Delta H_{TS-D(T)} = 0$, the Massieu-Planck activation potential for folding ( $\Delta G_{TS-D(T)}/T$ ) is a minimum, and $\Delta G_{TS-D(T)}$ is purely entropic. The values of the reference temperatures are given in **Table 1**.

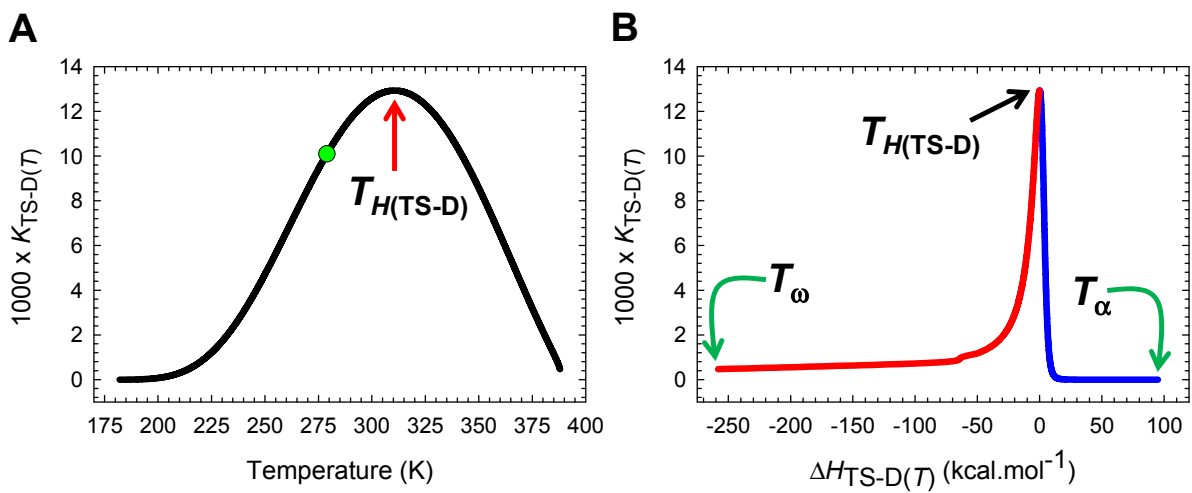

**Figure 1−figure supplement 2.**

**The solubility of the TSE relative to the DSE across a broad temperature regime.**

**(A)** $K_{TS\text{-}D(T)}$ is a maximum when $\Delta H_{TS\text{-}D(T)} = 0$, or when the Massieu-Planck activation potential for folding, $\Delta G_{TS\text{-}D(T)}/T$, is a minimum, and this occurs precisely at $T_{H(TS\text{-}D)}$. The slope of this curve is given by $K_{TS\text{-}D(T)}\Delta H_{TS\text{-}D(T)}/RT^2$. The green dot represents $T_S$, the temperature at which $\Delta G_{D\text{-}N(T)}$ is a maximum, $\Delta G_{TS\text{-}D(T)}$ is a minimum, and the absolute entropies of the DSE, the TSE and the NSE are identical. **(B)** The solubility of the TSE as compared to the DSE is the greatest when $\Delta H_{TS\text{-}D(T)} = 0$, or equivalently, when the Gibbs barrier to folding is purely entropic. The slope of this curve is given by $K_{TS\text{-}D(T)}\Delta H_{TS\text{-}D(T)}/\Delta C_{pTS\text{-}D(T)}RT^2$. The blue and red sections of the curve represent the temperature regimes $T_\alpha \leq T \leq T_{H(TS\text{-}D)}$ and $T_{H(TS\text{-}D)} \leq T \leq T_\omega$, respectively. The values of the reference temperatures are given in **Table 1**.

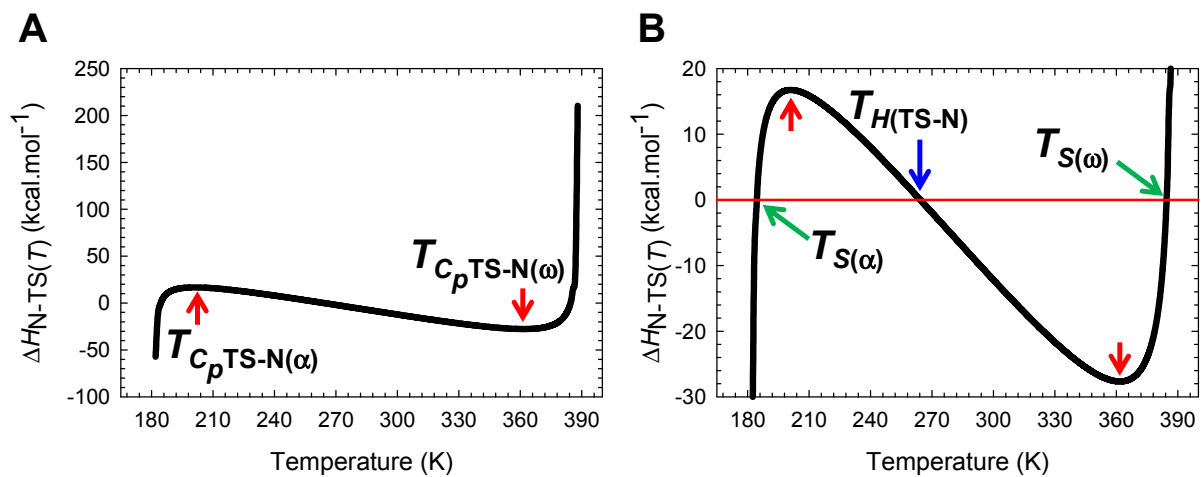

**Figure 2.**

**Temperature-dependence of the change in enthalpy for the partial folding reaction $[TS] \rightleftharpoons N$.**

**(A)** The variation in $\Delta H_{\text{N-TS}(T)}$ with temperature. The slope of this curve varies with temperature and equals $\Delta C_{p\text{N-TS}(T)}$. The red pointers indicate the temperatures where $\Delta C_{p\text{N-TS}(T)}$ (or $-\Delta C_{p\text{TS-N}(T)}$) is zero. **(B)** An appropriately scaled version of the plot on the left to illuminate the various temperature regimes. The net flux of the conformers from the TSE to the NSE is enthalpically: (*i*) favourable for $T_\alpha \leq T < T_{S(\alpha)}$ and $T_{H(\text{TS-N})} < T < T_{S(\omega)}$ ($\Delta H_{\text{N-TS}(T)} < 0$); (*ii*) unfavourable for $T_{S(\alpha)} < T < T_{H(\text{TS-N})}$ and $T_{S(\omega)} < T \leq T_\omega$ ($\Delta H_{\text{N-TS}(T)} > 0$); and (*iii*) neither favourable nor unfavourable at $T_{S(\alpha)}$, $T_{H(\text{TS-N})}$, and $T_{S(\omega)}$. At $T_{S(\alpha)}$ and $T_{S(\omega)}$, we have the unique scenario: $\Delta G_{\text{N-TS}(T)} = \Delta S_{\text{N-TS}(T)} = \Delta H_{\text{N-TS}(T)} = 0$, and $k_{u(T)} = k^0$. The values of the reference temperatures are given in **Table 1**.

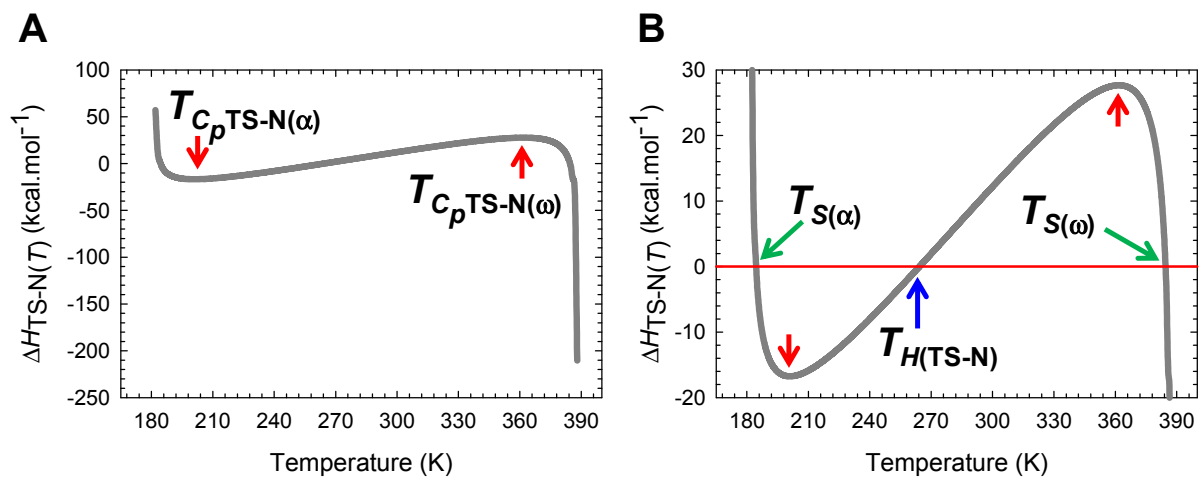

**Figure 2−figure supplement 1.**

**Temperature-dependence of the activation enthalpy for unfolding.**

**(A)** The variation in $\Delta H_{\text{TS-N}(T)}$ function with temperature. The slope of this curve equals $\Delta C_{p\text{TS-N}(T)}$ and is zero at $T_{C_p\text{TS-N}(\alpha)}$ and $T_{C_p\text{TS-N}(\omega)}$. **(B)** An appropriately scaled version of the figure on the left to illuminate the various temperature-regimes: (*i*) $\Delta H_{\text{TS-N}(T)} > 0$ for $T_\alpha \leq T < T_{S(\alpha)}$ and $T_{H(\text{TS-N})} < T < T_{S(\omega)}$; (*ii*) $\Delta H_{\text{TS-N}(T)} < 0$ for $T_{S(\alpha)} < T < T_{H(\text{TS-N})}$ and $T_{S(\omega)} < T \leq T_\omega$; and (*iii*) $\Delta H_{\text{TS-N}(T)} = 0$ at $T_{S(\alpha)}$, $T_{H(\text{TS-N})}$, and $T_{S(\omega)}$. Note that $k_{u(T)}$ is a minimum at $T_{H(\text{TS-N})}$. Further, at $T_{S(\alpha)}$ and $T_{S(\omega)}$, we have the unique scenario: $\Delta G_{\text{TS-N}(T)} = \Delta S_{\text{TS-N}(T)} = \Delta H_{\text{TS-N}(T)} = 0$, and $k_{u(T)} = k^0$, i.e., unfolding is barrierless; and for the temperature regimes $T_\alpha \leq T < T_{S(\alpha)}$ and $T_{S(\omega)} < T \leq T_\omega$, unfolding is once again barrier-limited but falls under the *Marcus-inverted-regime*. This is in contrast to the *conventional barrier-limited* unfolding that occurs in the regime $T_{S(\alpha)} < T < T_{S(\omega)}$ (see **Figure 2−figure supplement 2**). The values of the reference temperatures are given in **Table 1**.

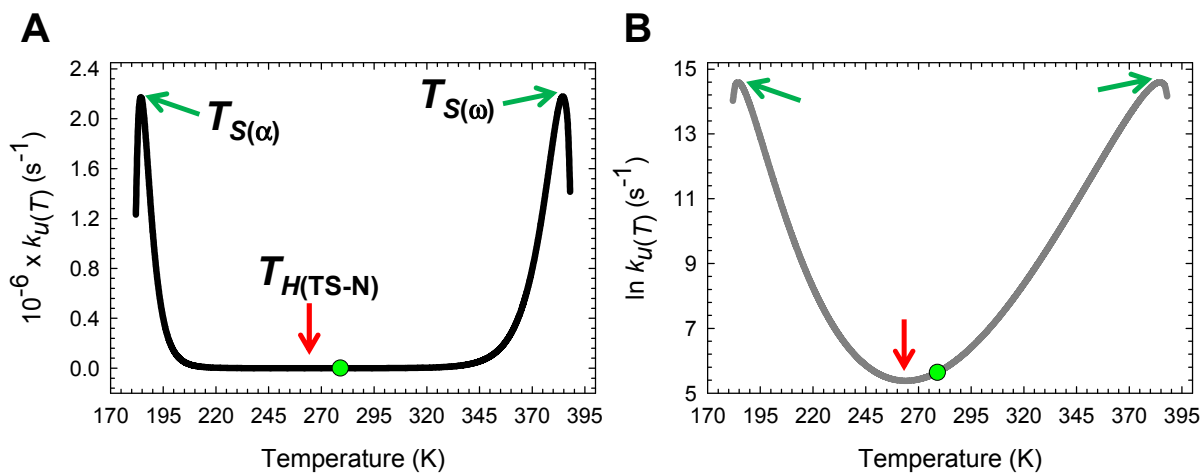

**Figure 2−figure supplement 2.**

**Temperature-dependence of $k_{u(T)}$.**

**(A)** Temperature-dependence of $k_{u(T)}$ on a linear scale. The slope of this curve is given by $k_{u(T)} \Delta H_{\text{TS-N}(T)} / RT^2$. Unlike $k_{f(T)}$ which has only one extremum, $k_{u(T)}$ is a minimum at $T_{H(\text{TS-N})}$ and a maximum ($k_{u(T)} = k^0$) at $T_{S(\alpha)}$ and $T_{S(\omega)}$ (green pointers). Although the minimum of $k_{u(T)}$ is not apparent on a linear scale, the *barrierless* and the *Marcus-inverted-regimes* for unfolding are readily apparent (see Paper-III).[3] **(B)** Temperature-dependence of $k_{u(T)}$ on a logarithmic scale. The slope of this curve is given by $\Delta H_{\text{TS-N}(T)} / RT^2$. The green dots represent the temperature $T_S$ at which $\Delta G_{\text{D-N}(T)}$ and $\Delta G_{\text{TS-N}(T)}$ are both a maximum, and the absolute entropies of the DSE, the TSE and the NSE are identical. The red pointers indicate the temperature $T_{H(\text{TS-N})}$ at which $k_{u(T)}$ is a minimum, $\Delta H_{\text{TS-N}(T)} = 0$, the Massieu-Planck activation potential for unfolding ($\Delta G_{\text{TS-N}(T)} / T$) is a maximum, and $\Delta G_{\text{TS-N}(T)}$ is purely entropic. The values of the reference temperatures are given in **Table 1**.

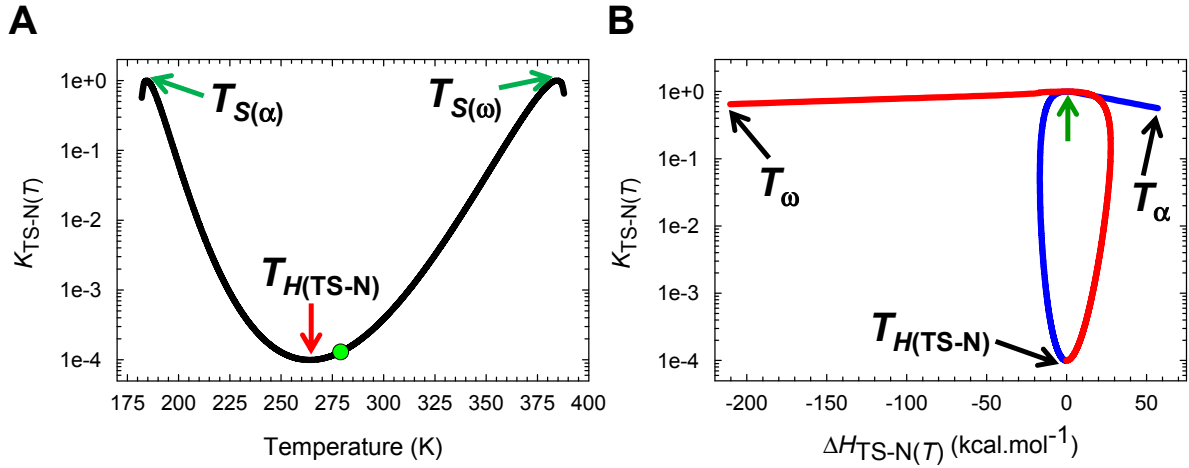

**Figure 2−figure supplement 3.**

**The solubility of the TSE relative to the NSE across a broad temperature regime.**

**(A)** $K_{TS\text{-}N(T)}$ is a minimum when $\Delta H_{TS\text{-}N(T)} = 0$, or when the Massieu-Planck activation potential for unfolding, $\Delta G_{TS\text{-}N(T)}/T$, is a maximum; and this occurs precisely at $T_{H(TS\text{-}N)}$. Further, $K_{TS\text{-}N(T)}$ is unity at $T_{S(\alpha)}$ and $T_{S(\omega)}$. The slope of this curve is given by $K_{TS\text{-}N(T)} \Delta H_{TS\text{-}N(T)}/RT^2$. The ordinate is shown on a log scale (base 10) to illuminate the minimum of $K_{TS\text{-}N(T)}$. The green dot represents the temperature $T_S$ at which $\Delta G_{D\text{-}N(T)}$ and $\Delta G_{TS\text{-}N(T)}$ are both a maximum, and the absolute entropies of the DSE, the TSE and the NSE are identical. **(B)** The solubility of the TSE as compared to the NSE is the least when $\Delta H_{TS\text{-}N(T)} = 0$ or when the Gibbs barrier to unfolding is purely entropic. The slope of this curve is given by $K_{TS\text{-}N(T)} \Delta H_{TS\text{-}N(T)} / \Delta C_{p\text{TS-}N(T)} RT^2$. The point where the solubility of the TSE is identical to that of the NSE is indicated by the unlabelled green pointer, and described earlier, occurs precisely at $T_{S(\alpha)}$ and $T_{S(\omega)}$. The blue and red sections of the curve represent the temperature regimes $T_\alpha \leq T \leq T_{H(TS\text{-}N)}$ and $T_{H(TS\text{-}N)} \leq T \leq T_\omega$, respectively. Note that the ordinate is on a log scale (base 10). The values of the reference temperatures are given in **Table 1**.

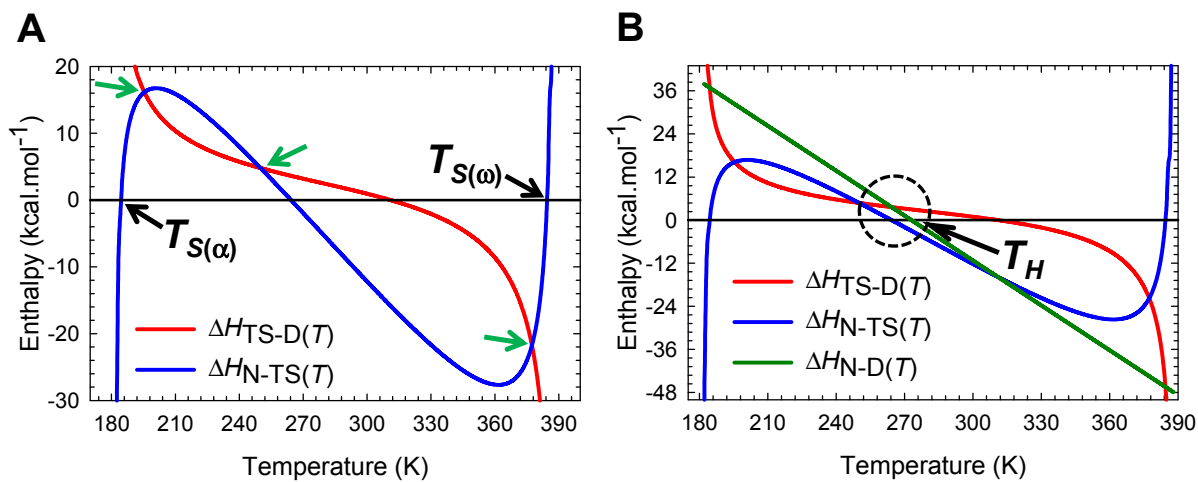

**Figure 3.**

**An overlay of $\Delta H_{TS\text{-}D(T)}$, $\Delta H_{N\text{-}TS(T)}$, and $\Delta H_{N\text{-}D(T)}$ functions.**

**(A)** An overlay of $\Delta H_{TS\text{-}D(T)}$ and $\Delta H_{N\text{-}TS(T)}$ functions. At the temperatures where the functions intersect (green pointers, 195.5 K, 250 K, and 377.4 K), the absolute enthalpy of the TSE is exactly half the algebraic sum of the absolute enthalpies of the DSE and the NSE, i.e., $H_{TS(T)} = \left(H_{D(T)} + H_{N(T)}\right)/2$. The intersection of the blue curve with the black reference line occurs at $T_{S(\alpha)}$, $T_{H(TS\text{-}N)}$, and $T_{S(\omega)}$. The intersection of the red curve with the black reference line occurs at $T_{H(TS\text{-}D)}$. **(B)** An overlay of $\Delta H_{TS\text{-}D(T)}$, $\Delta H_{N\text{-}TS(T)}$, and $\Delta H_{N\text{-}D(T)}$ functions to illuminate the relative contribution of the enthalpies of the partial folding reactions $D \rightleftharpoons [TS]$ and $[TS] \rightleftharpoons N$ to the change in enthalpy of folding at equilibrium. The red and the green curves intersect at $T_{S(\alpha)}$, $T_{H(TS\text{-}N)}$, and $T_{S(\omega)}$, and the blue and the green curves intersect at $T_{H(TS\text{-}D)}$. $\Delta H_{N\text{-}D(T)} = 0$ at the temperature ($T_H$) where the green curve intersects the black reference line. The net flux of the conformers from the DSE to the NSE is enthalpically unfavourable for $T_\alpha \leq T < T_H$, and favourable for $T_H < T \leq T_\omega$. The values of the reference temperatures are given in **Table 1**. See **Figure 3−figure supplement 1** for an appropriately scaled view of the intersections occurring inside the encircled region.

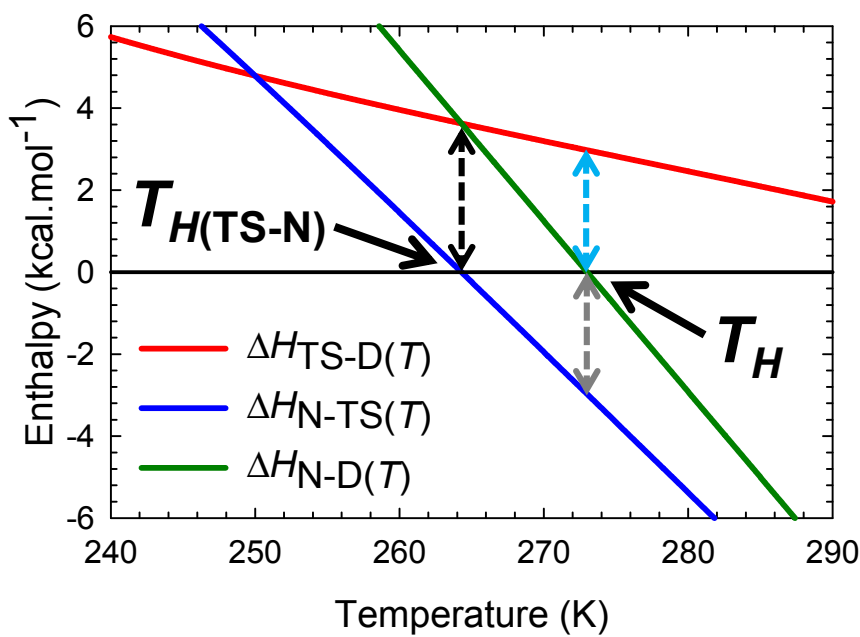

**Figure 3−figure supplement 1**

**An appropriately scaled view of the intersection of $\Delta H_{TS-D(T)}$, $\Delta H_{N-TS(T)}$, and $\Delta H_{N-D(T)}$ functions for the temperature regime $T_{H(TS-N)} < T < T_H$.**

Because $\Delta H_{TS-D(T)} > 0$ and $\Delta H_{N-TS(T)} < 0$ for $T_{H(TS-N)} < T < T_H$, the positive $\Delta H_{N-D(T)}$ that stems from the coupling of the partial folding reactions $D \rightleftharpoons [TS]$ and $[TS] \rightleftharpoons N$ is primarily due to the net heat released in $[TS] \rightleftharpoons N$ not fully compensating for the net heat absorbed to activate the denatured conformers to the TSE. The intersection of $\Delta H_{TS-D(T)}$, $\Delta H_{N-D(T)}$ functions (red and green curves) occurs precisely at $T_{H(TS-N)}$. When $T = T_H$, $\Delta H_{TS-D(T)} = -\Delta H_{N-TS(T)}$, i.e., the red and the blue curves are equidistant from the black horizontal reference line. The values of the reference temperatures are given in **Table 1**.

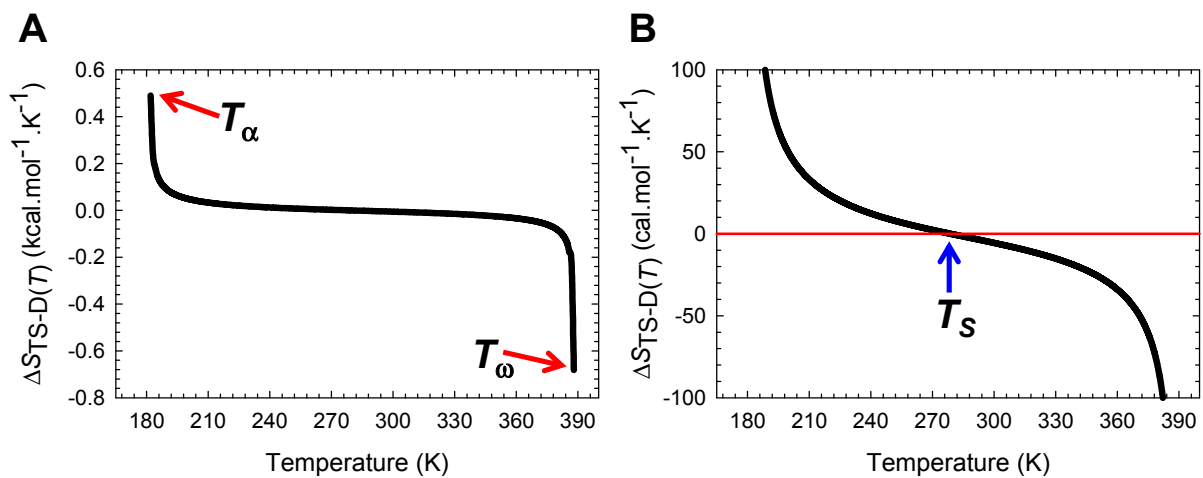

**Figure 4.**

**Temperature-dependence of the activation entropy for folding.**

**(A)** The variation in $\Delta S_{\text{TS-D}(T)}$ function with temperature. The slope of this curve varies with temperature and equals $\Delta C_{p\text{TS-D}(T)}/T$. **(B)** An appropriately scaled version of the figure on the left to illuminate the three temperature regimes and their implications: (*i*) $\Delta S_{\text{TS-D}(T)} > 0$ for $T_\alpha \leq T < T_S$; (*ii*) $\Delta S_{\text{TS-D}(T)} < 0$ for $T_S < T \leq T_\omega$; and (*iii*) $\Delta S_{\text{TS-D}(T)} = 0$ when $T = T_S$. The values of the reference temperatures are given in **Table 1**.

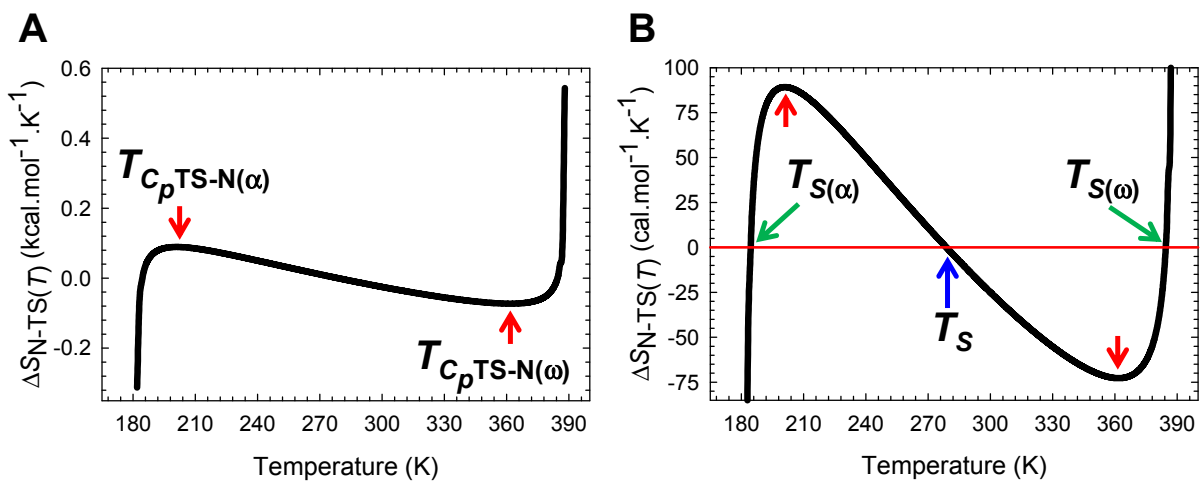

**Figure 5.**

**The variation in $\Delta S_{\text{N-TS}(T)}$ for the partial folding reaction $[TS] \rightleftharpoons N$.**

**(A)** The variation in $\Delta S_{\text{N-TS}(T)}$ function with temperature. The slope of this curve, given by $\Delta C_{p\text{N-TS}(T)}/T$, varies with temperature, and is zero at $T_{C_p\text{TS-N}(\alpha)}$ and $T_{C_p\text{TS-N}(\omega)}$. **(B)** An appropriately scaled view of the plot on the left to illuminate the various temperature regimes. The flux of the conformers from the TSE to the NSE is entropically: (*i*) unfavourable for $T_\alpha \leq T < T_{S(\alpha)}$ and $T_S < T < T_{S(\omega)}$ ($\Delta S_{\text{N-TS}(T)} < 0$); (*ii*) favourable for $T_{S(\alpha)} < T < T_S$ and $T_{S(\omega)} < T \leq T_\omega$ ($\Delta S_{\text{N-TS}(T)} > 0$); and (*iii*) neutral at $T_{S(\alpha)}$, $T_S$, and $T_{S(\omega)}$. The values of the reference temperatures are given in **Table 1**.

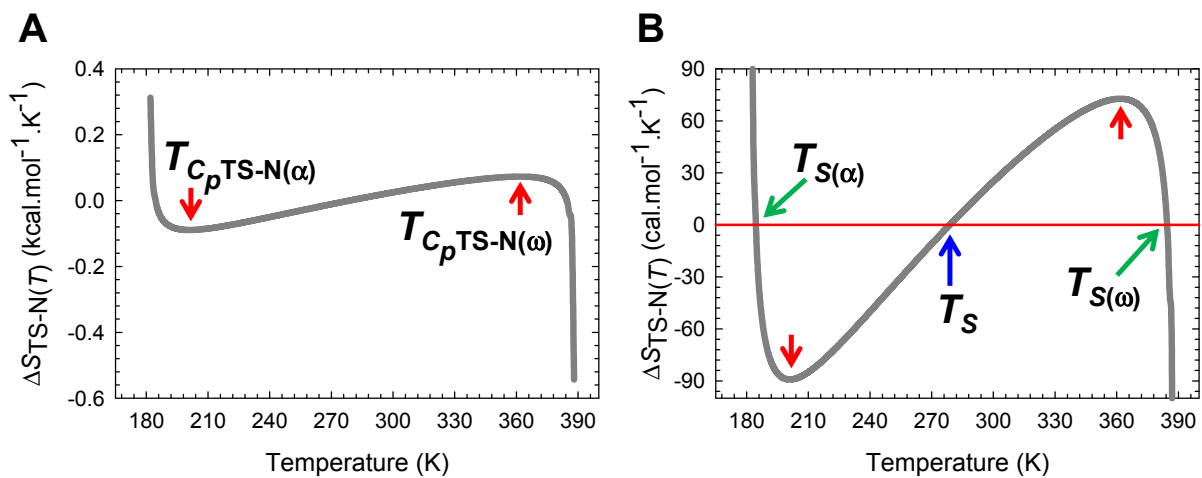

**Figure 5−figure supplement 1**

**Temperature-dependence of the activation entropy for unfolding.**

(A) The variation in $\Delta S_{\text{TS-N}(T)}$ function with temperature for the partial unfolding reaction $N \rightleftharpoons [TS]$. The slope of this curve, given by $\Delta C_{p\text{TS-N}(T)}/T$, and is zero at $T_{C_p\text{TS-N}(\alpha)}$ and $T_{C_p\text{TS-N}(\omega)}$. (B) An appropriately scaled version of the figure on the left to illuminate the temperature regimes and their implications: (*i*) $\Delta S_{\text{TS-N}(T)} > 0$ for $T_\alpha \leq T < T_{S(\alpha)}$ and $T_S < T < T_{S(\omega)}$; (*ii*) $\Delta S_{\text{TS-N}(T)} < 0$ for $T_{S(\alpha)} < T < T_S$ and $T_{S(\omega)} < T \leq T_\omega$; and (*iii*) $\Delta S_{\text{TS-N}(T)} = 0$ at $T_{S(\alpha)}$, $T_S$, and $T_{S(\omega)}$. Note that at $T_{S(\alpha)}$ and $T_{S(\omega)}$, we have the unique scenario: $\Delta G_{\text{TS-N}(T)} = \Delta S_{\text{TS-N}(T)} = \Delta H_{\text{TS-N}(T)} = 0$, and $k_{u(T)} = k^0$, i.e., unfolding is barrierless; and for the temperature regimes $T_\alpha \leq T < T_{S(\alpha)}$ and $T_{S(\omega)} < T \leq T_\omega$, unfolding is once again barrier-limited but falls under the *Marcus-inverted-regime*. This is in contrast to the *conventional barrier-limited* unfolding that occurs in the regime $T_{S(\alpha)} < T < T_{S(\omega)}$. The values of the reference temperatures are given in **Table 1**.

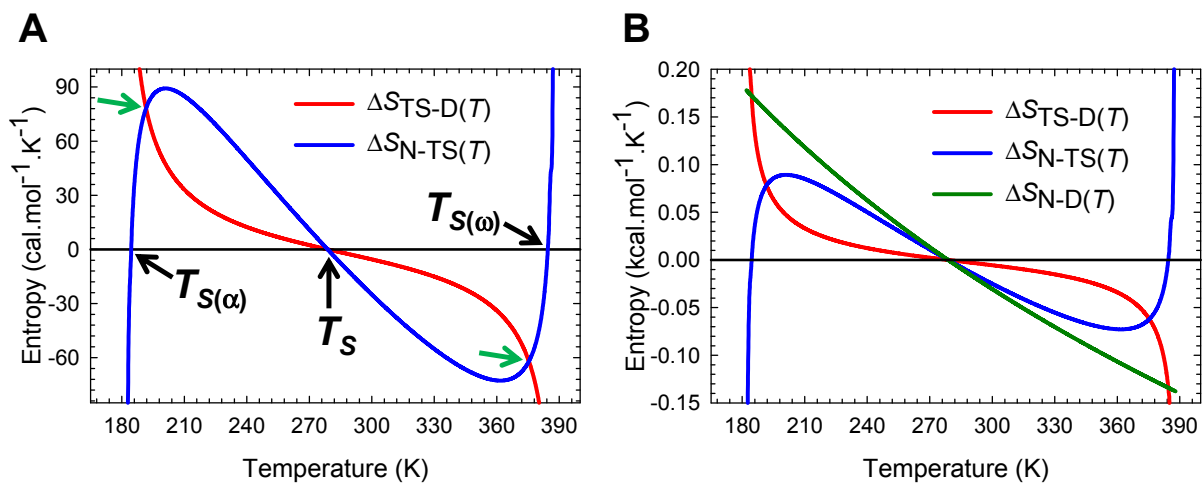

**Figure 6.**

**An overlay of $\Delta S_{TS-D(T)}$, $\Delta S_{N-TS(T)}$, and $\Delta S_{N-D(T)}$ functions.**

**(A)** An overlay of $\Delta S_{TS-D(T)}$ and $\Delta S_{N-TS(T)}$ functions for the partial folding reactions $D \rightleftharpoons [TS]$ and $[TS] \rightleftharpoons N$, respectively. At the high and low temperatures where the functions intersect (green pointers, 191.7 K and 375.5 K), the absolute entropy of the TSE is exactly half the algebraic sum of the absolute entropies of the DSE and the NSE, i.e., $S_{TS(T)} = \left( S_{D(T)} + S_{N(T)} \right)/2$. The intersection of the blue curve with the black reference line occurs at $T_{S(\alpha)}$, $T_S$, and $T_{S(\omega)}$. The intersection of the red curve with the black reference line occurs at $T_S$. **(B)** An overlay of $\Delta S_{TS-D(T)}$, $\Delta S_{N-TS(T)}$, and $\Delta S_{N-D(T)}$ functions to illuminate the relative contribution of the entropies of the partial folding reactions $D \rightleftharpoons [TS]$ and $[TS] \rightleftharpoons N$ to the change in entropy for the coupled reaction $D \rightleftharpoons N$. While the red and the green curves intersect at $T_{S(\alpha)}$, $T_S$, and $T_{S(\omega)}$, all the three curves intersect at $T_S$ ( $S_{D(T)} = S_{TS(T)} = S_{N(T)}$ ). The net flux of the conformers from the DSE to the NSE is entropically favourable for $T_\alpha \leq T < T_S$ and unfavourable for $T_S < T \leq T_\omega$. The values of the reference temperatures are given in **Table 1**.

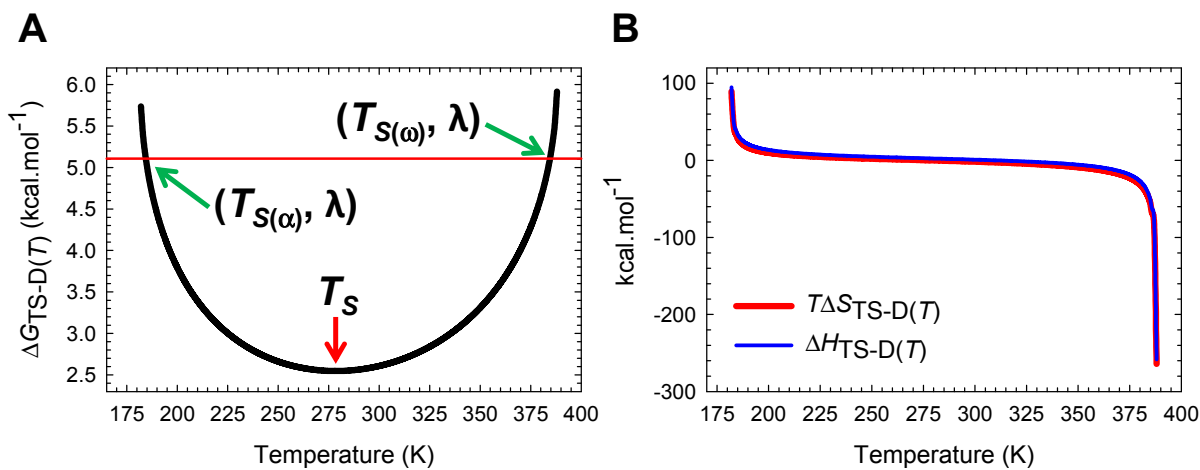

**Figure 7.**

**Temperature-dependence of the Gibbs activation energy for folding.**

**(A)** $\Delta G_{TS\text{-}D(T)}$ is a minimum at $T_S$, identical to $\lambda = \alpha \left( m_{D\text{-}N} \right)^2 = 5.106$ kcal.mol$^{-1}$ at $T_{S(\alpha)}$ and $T_{S(\omega)}$, and greater than $\lambda$ for $T_\alpha \leq T < T_{S(\alpha)}$ and $T_{S(\omega)} < T \leq T_\omega$ ($\lambda$ is the *Marcus reorganization energy* for protein folding), and $\partial \Delta G_{TS\text{-}D(T)} / \partial T = -\Delta S_{TS\text{-}D(T)} = 0$ at $T_S$. **(B)** Despite large changes in $\Delta H_{TS\text{-}D(T)}$ (~ 400 kcal.mol$^{-1}$) $\Delta G_{TS\text{-}D(T)}$ varies only by ~3.4 kcal.mol$^{-1}$ due to compensating changes in $\Delta S_{TS\text{-}D(T)}$. See the appropriately scaled figure supplement for description.

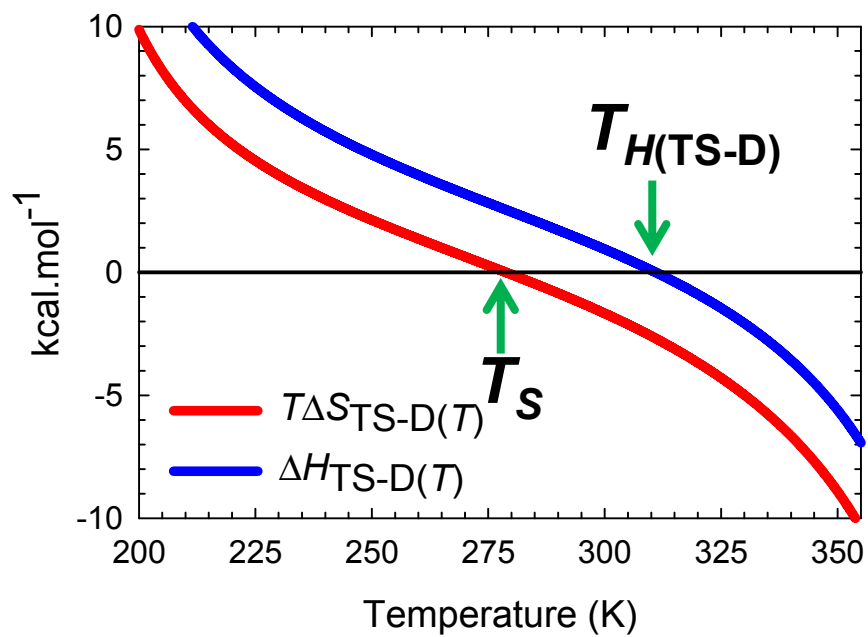

**Figure 7−figure supplement 1.**

**Deconvolution of the Gibbs activation energy for the reaction $D \rightleftharpoons [TS]$..**

For $T_\alpha \leq T < T_S$, $T\Delta S_{TS-D(T)} > 0$ but is more than offset by unfavourable $\Delta H_{TS-D(T)}$, leading to incomplete compensation and a positive $\Delta G_{TS-D(T)}$ ($\Delta H_{TS-D(T)} - T\Delta S_{TS-D(T)} > 0$). When $T = T_S$, $\Delta G_{TS-D(T)}$ is a minimum and purely enthalpic ($\Delta G_{TS-D(T)} = \Delta H_{TS-D(T)} > 0$). For $T_S < T < T_{H(TS-D)}$, the activation is enthalpically and entropically disfavoured ($\Delta H_{TS-D(T)} > 0$ and $T\Delta S_{TS-D(T)} < 0$) leading to a positive $\Delta G_{TS-D(T)}$. In contrast, for $T_{H(TS-D)} < T \leq T_\omega$, $\Delta H_{TS-D(T)} < 0$ but is more than offset by the unfavourable entropy ($T\Delta S_{TS-D(T)} < 0$), leading once again to a positive $\Delta G_{TS-D(T)}$. When $T = T_{H(TS-D)}$, $\Delta G_{TS-D(T)}$ is purely entropic ($\Delta G_{TS-D(T)} = -T\Delta S_{TS-D(T)} > 0$) and $k_{f(T)}$ is a maximum.

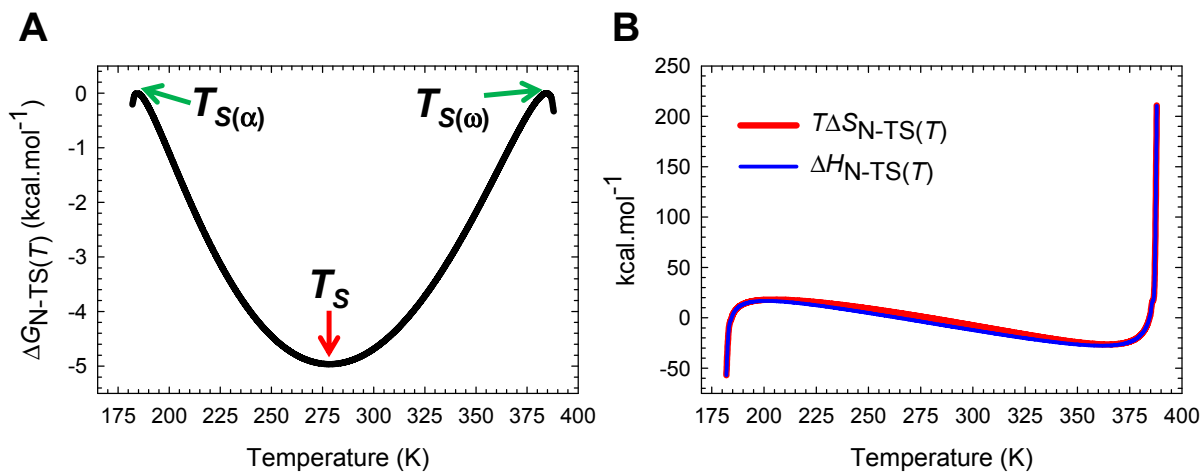

**Figure 8.**

**Temperature-dependence of the change in Gibbs energy for the partial folding reaction $[TS] \rightleftharpoons N$**

**(A)** In contrast to $\Delta G_{TS-D(T)}$ which has only one extremum, $\Delta G_{N-TS(T)}$ is a minimum at $T_S$ and a maximum (zero) at $T_{S(\alpha)}$ and $T_{S(\omega)}$; consequently, $\partial \Delta G_{N-TS(T)}/\partial T = -\Delta S_{N-TS(T)} = 0$ at $T_{S(\alpha)}$, $T_S$ and $T_{S(\omega)}$. The values of the reference temperatures are given in **Table 1**. **(B)** Despite large changes in $\Delta H_{N-TS(T)}$, $\Delta G_{N-TS(T)}$ varies only by ~5 kcal.mol$^{-1}$ due to compensating changes in $\Delta S_{N-TS(T)}$. See the appropriately scaled figure supplement for description.

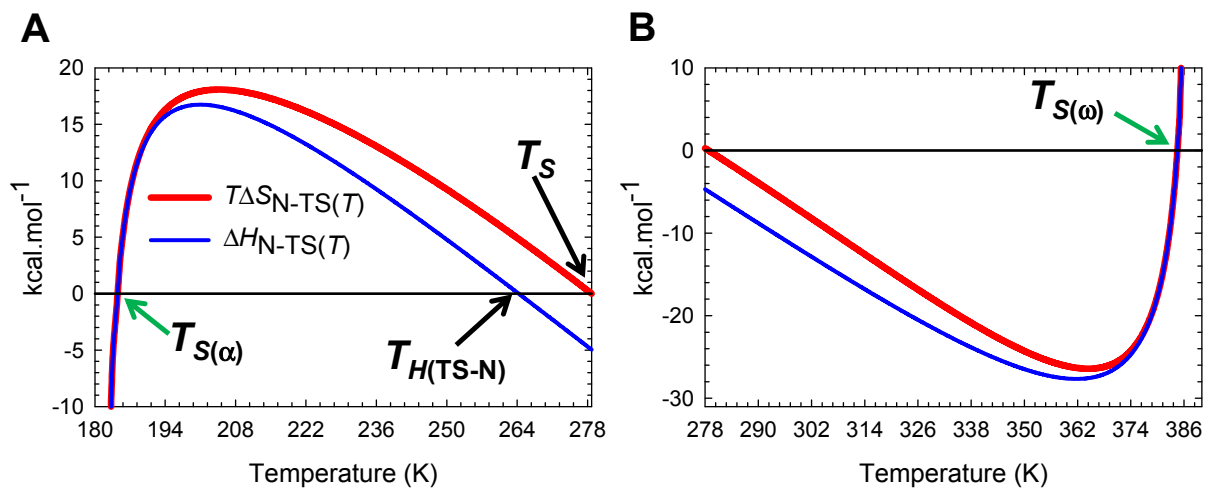

**Figure 8−figure supplement 1.**

**Deconvolution of the change in Gibbs energy for the partial folding reaction $[TS] \rightleftharpoons N$.**

These are appropriately scaled split views of **Figure 8B**. **(A)** For $T_\alpha \leq T < T_{S(\alpha)}$, $[TS] \rightleftharpoons N$ is entropically disfavoured ($T\Delta S_{\text{N-TS}(T)} < 0$) but is more than compensated by the exothermic enthalpy ($\Delta H_{\text{N-TS}(T)} < 0$), leading to $\Delta G_{\text{N-TS}(T)} < 0$. When $T = T_{S(\alpha)}$, $\Delta S_{\text{N-TS}(T)} = \Delta H_{\text{N-TS}(T)} = \Delta G_{\text{N-TS}(T)} = 0$, and the net flux of the conformers from the TSE to the NSE is zero. For $T_{S(\alpha)} < T < T_{H(\text{TS-N})}$, $[TS] \rightleftharpoons N$ is enthalpically unfavourable ($\Delta H_{\text{N-TS}(T)} > 0$) but is more than compensated by entropy ($T\Delta S_{\text{N-TS}(T)} > 0$) leading to $\Delta G_{\text{N-TS}(T)} < 0$. When $T = T_{H(\text{TS-N})}$, the net flux from the TSE to the NSE is driven purely by the favourable change in entropy ($\Delta G_{\text{N-TS}(T)} = -T\Delta S_{\text{N-TS}(T)} < 0$). For $T_{H(\text{TS-N})} < T < T_S$, the net flux of the conformers from the TSE to the NSE is entropically and enthalpically favourable ($\Delta H_{\text{N-TS}(T)} < 0$ and $T\Delta S_{\text{N-TS}(T)} > 0$) leading to $\Delta G_{\text{N-TS}(T)} < 0$. When $T = T_S$, the net flux is driven purely by the exothermic change in enthalpy ($\Delta G_{\text{N-TS}(T)} = \Delta H_{\text{N-TS}(T)} < 0$). **(B)** For $T_S < T < T_{S(\omega)}$, $[TS] \rightleftharpoons N$ is entropically unfavourable ($T\Delta S_{\text{N-TS}(T)} < 0$) but is more than compensated by the exothermic enthalpy ($\Delta H_{\text{N-TS}(T)} < 0$) leading to $\Delta G_{\text{N-TS}(T)} < 0$. When $T = T_{S(\omega)}$, $\Delta S_{\text{N-TS}(T)} = \Delta H_{\text{N-TS}(T)} = \Delta G_{\text{N-TS}(T)} = 0$, and the net flux of the conformers from the TSE to the NSE is zero. For $T_{S(\omega)} < T \leq T_\omega$, $[TS] \rightleftharpoons N$ is enthalpically unfavourable ($\Delta H_{\text{N-TS}(T)} > 0$) but is more than compensated by the favourable change in entropy ($T\Delta S_{\text{N-TS}(T)} > 0$), leading to $\Delta G_{\text{N-TS}(T)} < 0$.

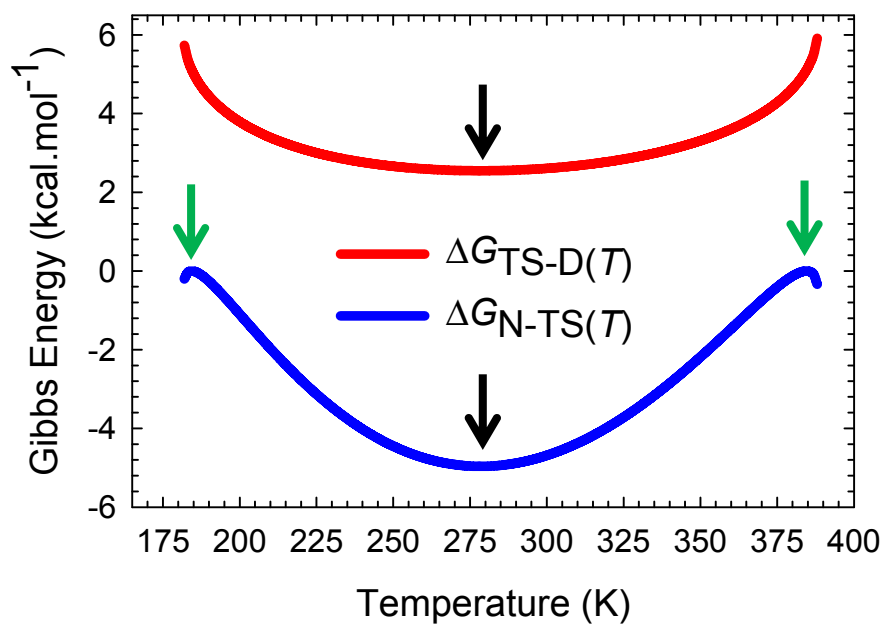

**Figure 9.**

**An overlay of $\Delta G_{TS\text{-}D(T)}$ and $\Delta G_{N\text{-}TS(T)}$ functions.**

Although both $\Delta G_{TS\text{-}D(T)}$ and $\Delta G_{N\text{-}TS(T)}$ are a minimum at $T_S$ (black pointers), $\Delta G_{TS\text{-}D(T)}$ is always positive and $\Delta G_{N\text{-}TS(T)}$ is negative except for the temperatures $T_{S(\alpha)}$ and $T_{S(\omega)}$ (green pointers). Further, $\Delta G_{TS\text{-}D(T)}$ is identical to the *Marcus reorganization energy* at $T_{S(\alpha)}$ and $T_{S(\omega)}$ (see Paper-III).[3]

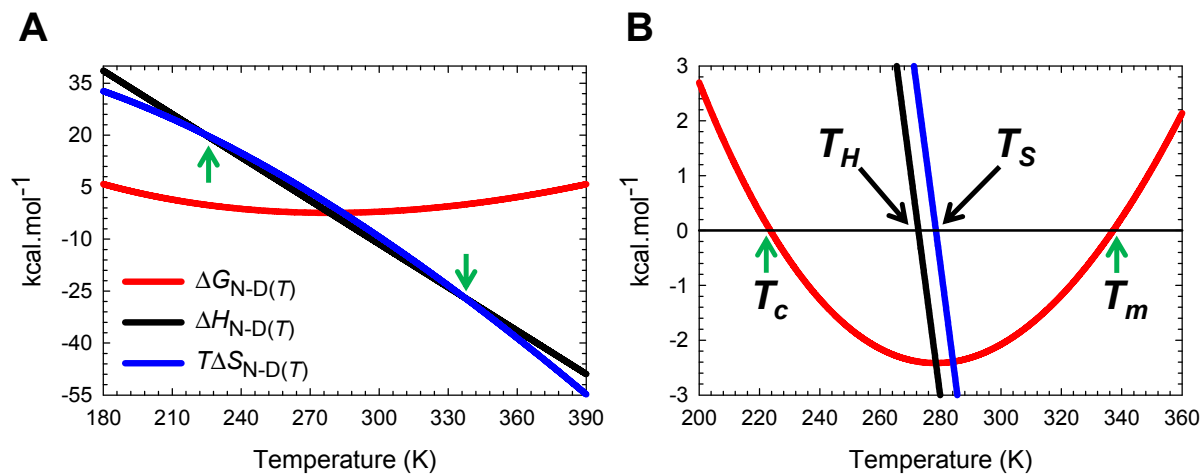

**Figure 10.**

**Stability curve for the folding reaction $D \rightleftharpoons N$.**

(A) Temperature-dependence of $\Delta G_{\text{N-D}(T)}$, $\Delta H_{\text{N-D}(T)}$, and $T\Delta S_{\text{N-D}(T)}$. The green pointers identify the cold ($T_c$) and heat ($T_m$) denaturation temperatures. The green pointers identify $T_c$ and $T_m$. The slopes of the red and black curves are given by $\partial \Delta G_{\text{N-D}(T)}/\partial T = -\Delta S_{\text{N-D}(T)}$ and $\partial \Delta H_{\text{N-D}(T)}/\partial T = \Delta C_{p\,\text{N-D}}$, respectively. (B) An appropriately scaled version of plot on the left. $T_H$ is the temperature at which $\Delta H_{\text{N-D}(T)} = 0$, and $T_S$ is the temperature at which $\Delta S_{\text{N-D}(T)} = 0$. The values of the reference temperatures are given in **Table 1**.

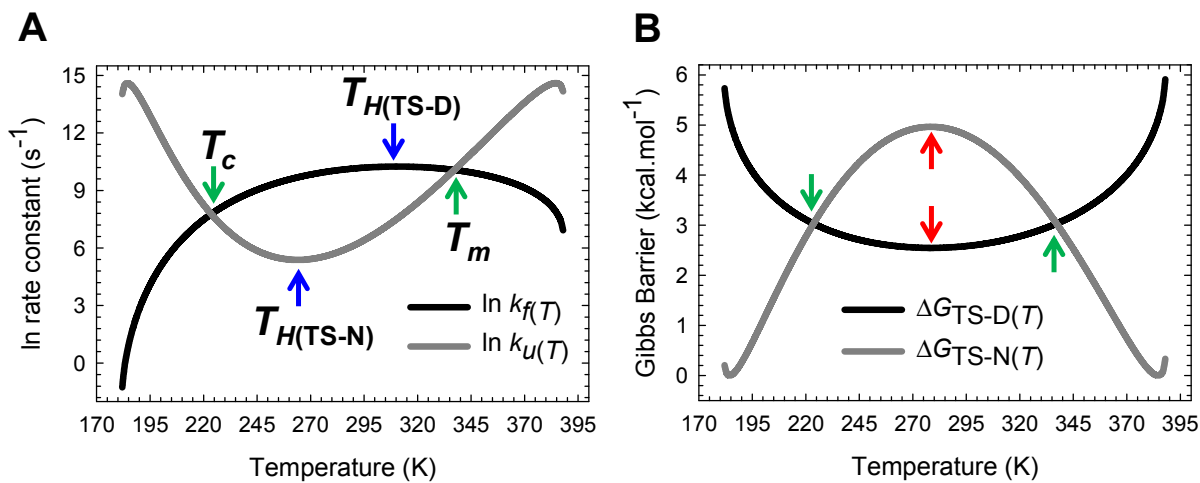

**Figure 10−figure supplement 1.**

**The principle of microscopic reversibility.**

**(A)** $k_{f(T)}$ is a maximum at $T_{H(TS-D)}$ and $k_{u(T)}$ is a minimum at $T_{H(TS-N)}$. The slopes of the black and grey curves are given by $\Delta H_{TS-D(T)}/RT^2$ and $\Delta H_{TS-N(T)}/RT^2$, respectively. **(B)** $\Delta G_{TS-D(T)}$ and $\Delta G_{TS-N(T)}$ are a minimum and a maximum, respectively, at $T_S$ (red pointers) leading to $\Delta G_{D-N(T)}$ being a maximum (or $\Delta G_{N-D(T)}$ a minimum) at $T_S$. Equilibrium stability is thus a consequence or the equilibrium manifestation of the underlying kinetic behaviour. The rate constants are identical at $T_c$ and $T_m$, leading to $\Delta G_{D-N(T)} = RT \ln\left(k_{f(T)}/k_{u(T)}\right) = \Delta G_{TS-N(T)} - \Delta G_{TS-D(T)} = 0$.

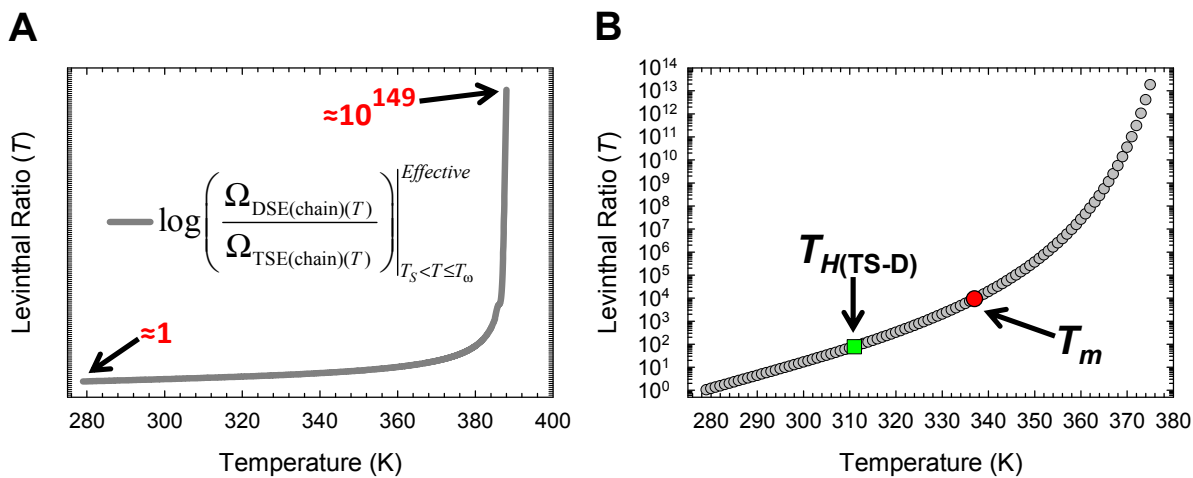

**Figure 11.**

**The ratio of effective number of accessible chain conformations in the DSE to those in the TSE.**

**(A)** Temperature-dependence of the change in the ratio of the effective number of accessible conformations in the DSE to those in the TSE calculated according to Eq. (55) and shown on a log scale (base 10). **(B)** An appropriately scaled version of the plot on the left. Although the effective ratio is greater than the Avogadro number for $T > \sim 382$ K, and is about $10^{149}$ when $T = T_\omega$, it is reasonably small for the temperature regime $T_S < T < T_m$. These calculations cannot be performed for $T < T_S$ since the entropy of solvent-release more than compensates for the decrease in chain entropy (**Figure 4B**).